\documentclass[aps,pre,twocolumn,showpacs,superscriptaddress]{revtex4-2}

\usepackage{amsmath}
\usepackage{amssymb}
\usepackage{graphicx}
  \usepackage{hyperref}
\usepackage[arrow, matrix, curve]{xy}
\usepackage{bm}
\usepackage{yhmath}
\usepackage{color}
\usepackage{url}
\newcommand\diff{\mathrm{d}}

\DeclareMathOperator\Imag{Im}

\renewcommand\vec[1]{\boldsymbol{\mathrm{#1}}}

\usepackage[usenames,dvipsnames]{xcolor}
\hypersetup{colorlinks=true, linkcolor=BrickRed, urlcolor=blue!50!black, citecolor=blue!50!black}

\usepackage[normalem]{ulem}

\makeatletter
\newcommand\hide@visible[1]{%
  \bgroup\fboxsep=.3ex\colorbox{Gray}{begin hide}%
  #1\colorbox{Gray}{end hide}\egroup%
}
\newcommand\hide@hidden[1]{%
  \bgroup\fboxsep=.3ex\colorbox{Gray}{hidden text}%
}
\newcommand\hide@invisible[1]{}
\newcommand\makevisible{\let\hide\hide@visible}
\newcommand\makehidden{\let\hide\hide@hidden}
\newcommand\makeinvisible{\let\hide\hide@invisible}
\makeatother
\makehidden

\begin{document}

\texttt{Published: Physical Review E 112, 015405 (2025)}
\title{Mode-coupling theory of the glass transition for a liquid in a    periodic potential}

\author{Abolfazl Ahmadirahmat}
\affiliation{Institut f{\"u}r Theoretische Physik,  Universit\"at Innsbruck, Technikerstra{\ss}e 21A, A-6020 Innsbruck, Austria}

\author{Michele Caraglio}
\affiliation{Institut f{\"u}r Theoretische Physik,  Universit\"at Innsbruck,
Technikerstra{\ss}e 21A, A-6020 Innsbruck, Austria}
\author{Vincent Krakoviack}
\affiliation{{\'E}cole Normale Sup{\'e}rieure de Lyon, CNRS, Laboratoire de Chimie UMR 5182 and Centre Blaise Pascal,  46 All{\'e}e d'Italie, 69364 Lyon, France}

\author{Thomas Franosch}
\affiliation{Institut f{\"u}r Theoretische Physik, Universit\"at Innsbruck,
Technikerstra{\ss}e 21A, A-6020 Innsbruck, Austria}

\email[]{thomas.franosch@uibk.ac.at}

\date{\today}

\begin{abstract}
We derive a microscopic theory for the structural dynamics in the vicinity of the glass transition for a liquid exposed to  a one-dimensional periodic potential. 
The periodic potential breaks translational invariance, in particular, the density  exhibits a periodic modulation. Using techniques familiar from solid-state theory, we define generalized intermediate scattering  functions from fluctuating densities in wave-vector space. Exact equations of motion are derived within the Mori-Zwanzig projection-operator formalism reflecting the residual lattice symmetries. Due to the lack of rotational symmetry it is necessary to split the currents into components parallel and perpendicular to the  modulation. We provide a closure of the equations in terms of a mode-coupling approximation for the force kernel. The theory reflects the usual analytic properties of correlation functions and encodes all phenomena known for mode-coupling theories.  We prove that the theory reduces to the conventional mode-coupling theory in the case of vanishing amplitude of the modulation. 
\end{abstract}

\maketitle

\section{Introduction} 
In many liquids the transition to a crystal can be circumvented by rapid cooling or compression leading to a metastable supercooled  state where transport coefficients such as the viscosity or diffusion coefficients vary by orders of magnitude upon small changes of temperature or density~\cite{Goetze:Complex_Dynamics}.  Once the structural relaxation time becomes macroscopic a glassy state is reached characterized by a  structural arrest concomitant  with the emergence of an amorphous elastic solid.  The slowing down of transport is accompanied  by a plethora of striking phenomena, such as the time-temperature superposition principle, the universal $\beta$-relaxation window, as well as the emergence of a two-time fractal. Many of the features have been rationalized within the mode-coupling theory of the glass transition (MCT)~\cite{Goetze:Complex_Dynamics,Janssen:2018}, a microscopic theory for the directly measurable intermediate scattering functions. 
Although MCT is considered as incomplete    since it misses certain relevant features, such as the ergodicity restoring processes, it serves as a  benchmark for various complementary approaches.

For a better understanding of the merits and limitations of the MCT approximation, the theory has been explored in multiple directions by adding more experimentally accessible control parameters, going beyond simple one-component systems as well as  considering situations where the liquid does not display all symmetries of a bulk liquid. For example,
 mixing polymers and colloids introduces an effective short-range attraction leading to competing mechanisms for forming a glass, which results in a complex nonequilibrium state diagram characterized by higher-order singularities and reentrant behavior
\cite{Bergenholtz:PRE_59:1999,Fabbian:PRE_59:1999,Dawson:PRE_63:2000,Pham:Science_296:2002}.  
Confining the liquid to disordered pores introduces the packing fraction of the quenched matrix as additional control parameter and yields  nontrivial transition scenarios, including reentrant and higher-order singularities  
\cite{Krakoviack:PRL_94:2005, *Krakoviack:PRE_75:2007, *Krakoviack:PRE_79:2009,Kurzidem:PRL_103:2009,Kim:EPL_88:2009}. 
Reentrant behavior also emerges upon confining simple liquids to a slit due to the interplay of  layering and local packing~\cite{Lang:PRL_105:2010,Lang:PRE_86:2012,Mandal:NatComm_5:2014,Jung:PRE_102:2020, Schrack:JSTATMECH:2020}. In these confined systems the dynamics becomes anisotropic which suggested  splitting the currents into components parallel and perpendicular to the walls yielding a mathematical structure somewhat different from the conventional MCT. However, the main mathematical features of the theory, such as the analytic properties of the solutions~\cite{Goetze:JMath195:1995,Franosch:JStatPhys109:2002,Lang:JSTATMECH_12:2013} and  
the emergence of the $\beta$-scaling law~\cite{Jung:JSTATMECH:2020} can still be demonstrated. Splitting currents was also necessary  for the structural dynamics to be independent of kinetic parameters for 
 molecules with  additional orientational degrees of freedom~\cite{Franosch:PRE56:1997,Goetze:PRE61:2000,Scheidsteger:PRE56:1997, Kaemmerer:PRE58:1998,  *Kaemmerer:PRE58:1998a}.  Beyond equilibrium dynamics,  MCT has been tested successfully also for systems driven  by shear \cite{Fuchs:PRL89:2002,Brader:PRL98:2007, Brader:PRL101:2008}, by active rheology~\cite{Gazuz:PRL102:2009,Senbil:PRL122:2019,Gruber:PRE101:2020}, for driven granular gases~\cite{Kranz:PRL104:2010,Kranz:PRL121:2018,Kranz:PRF5:2020},
or for active particles 
 \cite{Szamel:PRE91:2015,Liluashvili:PRE96:2017,Reichert:PRE104:2021,Debets:JCP157:2022}.  To investigate the quality of the factorization, 
 generalized-mode-coupling theory \cite{Mayer:PRL97:2006,Janssen:PRL115:2015,  Luo:JCP153:2020, Luo:PRL129:2022}
has been developed to postpone the factorization approximation to higher levels with the hope to achieve better quantitative agreement with simulations. 
 
 In this work,  we continue in this direction  
by exposing the liquid to a periodic external potential. For the case of colloids in suspensions such external potentials can be conveniently created using laser beams as has been introduced in  the pioneering works by Ashkin on optical tweezers \cite{Ashkin:PRL_24:1970,Ashkin:Science_210:1980,Ashkin:PNAS_94:1997}. Since then  the use of optical fields has opened the possibility to control mesosized particles on the micrometer scale
~\cite{Dholakia:NaturePhotonics_5:2011,Marago:NatureNanotechnology_8:2013}. 
Therefore colloidal suspensions are ideally suited to explore the material properties in strong external fields -- strong means here way beyond the linear response regime -- to push our understanding of condensed matter problems to frontiers that can be hardly reached in molecular systems. 
The colloid-light interaction is approximately conservative, i.e.\@ it can be well represented by an external potential proportional to the light intensity, while the optical scattering and absorption is often negligible~\cite{Grier:Nature_424:2004,Marago:NatureNanotechnology_8:2013}. The simplest non-trivial potential landscape then consists of a sinusoidal laser intensity which can be achieved experimentally by interference of two coherent laser beams~\cite{Smith:OptLett_6:1981,Chowdhury:PRL_55:1985}. The relative incident angle can be varied to tune the wavelength of the potential modulation, while the intensity determines their amplitude.

Here, we address the structural dynamics of a liquid in such  a periodic potential relying on the mode-coupling approach. Although our main application will be  experimental realizations on colloidal monolayers~ \cite{Dalle-Ferrier:SM_7:2011, Jenkins:JoP_40:2008,Capellmann:JCP_148:2018},
 we shall formulate the mode-coupling theory for a $d$-dimensional system and a one-dimensional modulation. Also with a little more effort, our approach can be extended to arbitrary periodic potentials. 
 Although we are ultimately interested in colloidal dynamics we formulate the theory for Newtonian dynamics first, anticipating that the slow structural dynamics in the vicinity of the glass transition is independent of the details of the underlying dynamics \cite{Goetze:Complex_Dynamics, Franosch:JMCS235:1998}. The current work focuses on the formal derivation of the mode-coupling equations, consequences for the nonequilibrium state diagram and multiple reentrant behavior are elaborated for the case of a monolayer of hard disks in the companion paper~\cite{Ahmadirahmat:PRE:2025}. 
 
 This work is organized as follows: In Sec.~\ref{Sec:Model} we present the underlying microscopic model, introduce the relevant quantities of interest reflecting the symmetries of the set-up, and formulate appropriately generalized intermediate scattering functions (ISF). In Sec.~\ref{Sec:Mori_Zwanzig} we employ the Mori-Zwanzig projection-operator formalism to derive formally exact equations of motion for the ISF in terms of suitable memory kernels. We account for the induced breaking of isotropy  by splitting the currents into components parallel and perpendicular to the modulation resulting in generalized force kernels. In Sec.~\ref{Sec:MCT} we use a mode-coupling approach to derive an approximate expression of the force kernels in terms of bilinear combinations of the ISF itself.  Various technical derivations are deferred to the appendices. 
For reference, in Sec.~\ref{Sec:Nonergodicity} we specialize the MCT equation to the long-time limits of the ISF. In Section~\ref{Sec:bulk_MCT} we show that the theory reduces to the standard MCT of the glass transition in the case of vanishing modulation. A summary and concluding remarks are provided
 in Sec.~\ref{Sec:Summary}.

\section{Model and observable quantities}\label{Sec:Model}
\subsection{Model set-up} 
We consider a simple $d$-dimensional liquid of  $N$ identical classical particles enclosed 
in a hypercube  of volume $V= L^d$ in the presence of an external periodic potential. The thermodynamic limit $N\to \infty, V\to \infty$ with fixed number density $n_{0}=N/V$ will be anticipated.

The collection of  positions and momenta of the particles are denoted by 
$\vec{x}^N=(\vec{x}_1,\ldots, \vec{x}_N)  \in V^N$ and $ \vec{p}^N = (\vec{p}_1,\ldots,\vec{p}_N) \in (\mathbb{R}^{d})^N$.
The dynamics of the positions and momenta are prescribed by Hamilton's equations driven by the Hamilton function
\begin{equation}
 \mathcal{H}(\vec{x}^N,\vec{p}^N) = \sum_{n=1}^N \frac{\vec{p}_n^2}{2m} + V(\vec{x}^N) + U(\vec{x}^N), 
\end{equation}
where $m$ denotes the mass of the particles. 
Here the mutual interaction between the particles is assumed to be pairwise additive
\begin{equation}
 V(\vec{x}^N) = \sum_{n=1}^{N-1} \sum_{m=n+1}^N \mathcal{V}(|\vec{x}_n-\vec{x}_{m}|),
\end{equation}
where the pair potential $\mathcal{V}(\cdot)$ depends only on the magnitude of the separation of the pair,  
such that linear momentum and  angular momentum is preserved. 
Additionally, a periodic modulation is imposed by the external potential
\begin{equation}
 U(\vec{x}^N) = \sum_{n=1}^N \mathcal{U}(z_n),
\end{equation}
where the single-particle potential $\mathcal{U}(\cdot)$ acts only on the $z$-coordinates of the particles and 
is assumed to be periodic  with period $a$: 
$\mathcal{U}(z) = \mathcal{U}(z+a)$  (with $L/a \in \mathbb{N}$). Generally we write for a spatial coordinate $\vec{x}= (\vec{r},z)$ where  $\vec{r}$ is the component perpendicular to the modulation and   $z$ along the modulation.

 Then all lattice-periodic function can be expanded in terms of suitably adapted Fourier modes, e.g. the external potential
\begin{equation}\label{eq:periodic}
 \mathcal{U}(z) =   \sum_{\mu\in \mathbb{Z} } \mathcal{U}_{\mu} \exp(-i Q_\mu z ),
\end{equation}
where the one-dimensional wavevectors are discrete $Q_\mu = 2\pi \mu/a, \mu \in \mathbb{Z}$. The associated Fourier coefficients are obtained as
\begin{align}\label{eq:def_Fourier}
 \mathcal{U}_{\mu} & = \frac{1}{a} \int_{0}^a   \mathcal{U}(z) \exp(i Q_\mu z) \diff  z  . 
\end{align}
For real $\mathcal{U}(z)$  in real space, $\mathcal{U}_\mu^* = \mathcal{U}_{-\mu}\in \mathbb{C}$.  For symmetric functions $\mathcal{U}(z) = \mathcal{U}(-z)$ the Fourier coefficients additionally fulfill $\mathcal{U}_\mu = \mathcal{U}_{-\mu}$ in particular $\mathcal{U}_\mu \in \mathbb{R}$. 

By the imposed external potential the liquid will no longer be translationally invariant nor isotropic. However, the Hamilton function remains invariant under permutations of the particles,  a time shift, and time reversal. Furthermore it is  translationally invariant by a uniform shift of the positions perpendicular to the modulation and invariant under a simultaneous  reflection of all spatial coordinates on a plane that includes the direction of  the modulation.   
For $d\geq 3$  a rotation $\mathcal{R}_z$ leaving the direction of  modulation fixed is also a symmetry transform of the Hamiltonian $\mathcal{H}(\vec{x}^N, \vec{p}^N)  = \mathcal{H}(\mathcal{R}_z \vec{x}^N, \vec{p}^N) $. Last, by discrete symmetry $\mathcal{H}$ remains unchanged by a uniform shift of all positions 
by integer multiples of the primitive vector $\vec{a} = a \vec{e}_z$. Together with the continuous translational symmetry perpendicular to the modulation, the Hamilton function is invariant under uniform shifts by vectors
$\vec{R}
\in \Lambda := \{\vec{r}+  n \vec{a}\in \mathbb{R}^d : \vec{r} \perp \vec{a},  n \in \mathbb{Z} \}$ of the degenerate Bravais lattice $\Lambda$. For symmetric potentials, $\mathcal{H}$ also remains invariant under  simultaneous reflection of all coordinates $(\vec{r}_n, z_n) \mapsto (\vec{r}_n, -z_n)$ along the modulation. Together with the reflection symmetries for planes including the direction of modulation this is equivalent to spatial inversion symmetry $\mathcal{H}(-\vec{x}^N, \vec{p}^N) = \mathcal{H}(\vec{x}^N,\vec{p}^N)$. We assume that the liquid reflects all these symmetries, in particular, there is no spontaneous symmetry breaking to a crystalline  or chiral phase, or  quasi long-range order in the positions or orientations.

\subsection{Observables} 

Here we are concerned with the dynamics of fluctuating densities
\begin{equation}
 \rho(\vec{x},t) := \sum_{n=1}^N \delta[\vec{x}-\vec{x}_n(t)] ,
\end{equation}
where $\vec{x}_l(t)$ denotes the position of particle $l$ at time $t$. 
We also define  the equilibrium density  
\begin{equation}
 n(z) := \langle \rho(\vec{x},t) \rangle ,
\end{equation}
where $\langle \cdot \rangle$ indicates canonical averaging. By the  translational symmetry of the Bravais lattice and in the liquid phase  it 
depends only on $z$ and is lattice-periodic: $n(z) = n(z+a)$. In particular it can be represented as in Eq.\eqref{eq:periodic} by its associated Fourier components $n_\mu\in \mathbb{C}, \mu \in \mathbb{Z}$. 
We shall also need the local volume, $v(z) := 1/n(z)$.  Then  the convolution theorem yields the relation for the Fourier coefficients
\begin{equation}\label{eq:convolution_theorem}
 \sum_{\kappa\in \mathbb{Z}} n_{\mu-\kappa}\, v_{\kappa-\nu} =  \delta_{\mu\nu} ,
\end{equation}
with the Kronecker symbol $\delta_{\mu\nu} = 1 $ if $\mu=\nu$ and 0 otherwise. Since the density and local volume in real space are non-negative $n(z) \geq 0, v(z) \geq 0$, one readily checks that the matrices $n_{\mu, \nu} := n_{\mu-\nu}, v_{\mu, \nu} := v_{\mu-\nu}$ are positive semidefinite. 

We then introduce fluctuations around the mean $\delta \rho(\vec{x},t) := \rho(\vec{x},t)-n(z)$ and define the density-density correlation function
\begin{equation}
 G(\vec{x},\vec{x}',t):=\frac{1}{n_{0}} \langle \delta\rho(\vec{x},t)\delta\rho(\vec{x}',0)\rangle,
\end{equation}
  also referred to as van Hove function~\cite{Hansen:Theory_of_Simple_Liquids}.
By translational and rotational symmetry perpendicular to the modulation, the van Hove function  
does not depend separately on $\vec{x}=(\vec{r},z), \vec{x}'= (\vec{r}', z')$, rather it depends only on the absolute distance of the perpendicular components $|\vec{r}-\vec{r}'|$ but explicitly on both parallel components $z, z'$.
 By time-reversal symmetry and time-translational symmetry 
it is an even function of time and symmetric with respect to interchanging the positions
\begin{equation}
 G(\vec{x}, \vec{x}',t) =  G(\vec{x},\vec{x}',-t)= G(\vec{x}',\vec{x},t).
\end{equation}
Furthermore,
 a common shift of the positions by a vector $\vec{R}\in \Lambda$ in the Bravais lattice leaves the van Hove function invariant
\begin{equation}
 G(\vec{x}+ \vec{R},\vec{x}'+\vec{R},t) =  G(\vec{x},\vec{x}',t).
\end{equation}
For a symmetric potential $G(\vec{x},\vec{x}',t)$ remains also invariant under simultaneous reflection $(\vec{r},z)  \mapsto ( \vec{r},-z),  (\vec{r}',z')  \mapsto  (\vec{r}',-z')$. 

Instead of relying on  real space we shall formulate the theory in wave-vector space. For a finite volume we need to introduce discrete Fourier transforms, a summary can be found in Appendix \ref{Sec:lattice_sums}.  Then the wave vectors $\vec{k}$ compatible with the finite volume $V$  are discrete in $\lambda^* = \{  \vec{k} \in \mathbb{R}^d: \vec{k} \in (2\pi \mathbb{Z}/L)^d \}$. 
We also introduce the reciprocal lattice $\Lambda^*$  associated with the modulation 
by requiring $\exp(i \vec{Q}_\mu \cdot \vec{R})= 1$ for all $\vec{Q}_\mu \in \Lambda^*, \vec{R} \in \Lambda$ which implies 
$\Lambda^* = \{ 
\vec{Q}_\mu = (2\pi \mu /a) \vec{e}_z : \mu \in \mathbb{Z} 
\} \subset \lambda^*$.  Any wave vector $\vec{k} \in \lambda^*$ can then be uniquely decomposed $\vec{k} = \vec{q} + \vec{Q}_\mu$ such that  $\vec{q} \in \text{BZ} := \{ \vec{q} \in \lambda^*: - \pi/a < \vec{q} \cdot \vec{e}_z \leq \pi/a \}$  is in the first Brillouin zone (BZ) and a reciprocal lattice vector $\vec{Q}_\mu \in \Lambda^*$. To simplify, we use the compact notation $\vec{q}_\mu := \vec{q} + \vec{Q}_\mu$. 
We define the fluctuating density to such a wave vector $\vec{q}_\mu$ as
\begin{align}
\delta \rho(\vec{q}_\mu,t) = &\delta \rho_\mu(\vec{q},t)  := \int_V \delta \rho(\vec{x},t) e^{i (\vec{q}+ \vec{Q}_\mu) \cdot \vec{x}} 
 \diff \vec{x} \nonumber  \\
&= 
\sum_{n=1}^N e^{i (\vec{q}+ \vec{Q}_\mu) \cdot \vec{x}_n(t) } 
 -V n_{\mu} \delta_{\vec{q},0} .
\end{align}
 Thus $\delta \rho_\mu(\vec{q},t)$ corresponds to the density modulation to wave vector $\vec{q}_\mu = \vec{q} + \vec{Q}_\mu$ and we use the notation $\delta \rho(\vec{q}_\mu,t) := \delta \rho_\mu(\vec{q},t)$ interchangeably. 
 We refer to Greek letters $\mu,\nu,\ldots$ as \emph{mode indices}.

Let us consider  the autocorrelation function $N^{-1} \langle \delta\rho_\mu(\vec{q},t)^* \delta \rho_\nu(\vec{k},0) \rangle$ with $\vec{q},\vec{k} \in \text{BZ}$, $\mu,\nu\in\mathbb{Z}$. Shifting all particle positions by  a lattice vector $\vec{R} \in \Lambda$ 
induces the mapping  $\delta \rho_\mu(\vec{q},t) \mapsto \exp(i \vec{q}\cdot \vec{R} )  \delta \rho_\mu(\vec{q},t)$.
Since this symmetry transform  has to leave the correlation function invariant, it is nonzero only for  $\vec{k}=\vec{q}$. 
We therefore define a properly generalized intermediate scattering function (ISF)
\begin{equation}
 S_{\mu\nu}(\vec{q},t) := \frac{1}{N} \langle \delta \rho_{\mu}(\vec{q},t)^* \delta \rho_{\nu}(\vec{q},0) \rangle,
\end{equation}
in particular, the initial values $S_{\mu\nu}(\vec{q}) := S_{\mu\nu}(\vec{q},t=0)$ will be referred to as generalized static structure factors. 
  The observation that $\vec{q} \in \text{BZ}$ has to be identical in both fluctuating densities is a manifestation of conservation of  crystal momentum. For $\mu=\nu$ we recover the conventional intermediate scattering function to wave vector $\vec{q}_\mu$. The novelty of the modulation is to explicitly allow for contributions $\mu\neq \nu$ such that the wave vectors differ by a reciprocal lattice vector $\vec{q}_\mu-\vec{q}_\nu \in \Lambda^*$.

By time inversion symmetry the ISF is even in time and by time-translation invariance it is a Hermitian  matrix
\begin{align}
S_{\mu\nu}(\vec{q},t) = S_{\mu\nu}(\vec{q},-t) = S_{\nu\mu}(\vec{q},t)^*. 
\end{align}
Moreover for every $\vec{q}\in \text{BZ}$ the ISF $S_{\mu\nu}(\vec{q},t)$ is a \emph{matrix-valued correlation function} 
with corresponding analytic properties \cite{Gesztesy:2000,Lang:JSTATMECH_12:2013}  since every contraction $\sum_{\mu,\nu \in \mathbb{Z}} y_\mu^* S_{\mu\nu}(\vec{q},t) y_\nu$ for arbitrary $y_\mu\in \mathbb{C}$ is merely the  autocorrelation function of $\sum_{\mu\in \mathbb{Z}} y_\mu \delta \rho_\mu(\vec{q},t)$. In particular, the generalized static structure factor $S_{\mu\nu}(\vec{q})$ is a positive semi-definite matrix. 

For $d\geq 3$ a rotation $\mathcal{R}_z$ leaving the axis of the modulation invariant is a symmetry of the ISF $S_{\mu\nu}(\vec{q},t) = S_{\mu\nu}(\mathcal{R}_z \vec{q},  t)$. Furthermore, reflecting $\vec{q}$ at any plane containing the direction of modulation leaves the ISF invariant, in particular, in 2D the wave vectors $\vec{q}= (q_x, q_z)$ and $(-q_x, q_z)$ yield the same ISF. 
For symmetric potentials spatial-inversion symmetry implies 
\begin{align}
S_{\mu\nu}(\vec{q},t) = S_{\mu\nu}(\vec{q},t)^* = S_{-\mu,-\nu}(-\vec{q},t), 
\end{align}
i.e. the matrices are real symmetric and invariant under a simultaneous change of sign of the mode indices. 

The respective inverse transformation for the fluctuating densities   reads (see Appendix \ref{Sec:lattice_sums})
\begin{align}
 \delta\rho(\vec{x},t) = \frac{1}{V} \sum_{\vec{q}\in \text{BZ} } \sum_{\mu\in\mathbb{Z}} \delta \rho_\mu(\vec{q},t) \exp[-\text{i} (\vec{q}+\vec{Q}_\mu)\cdot\vec{x} ] ,
\end{align}
and one readily checks that the ISF is obtained from the van Hove correlation function by spatial Fourier transform
\begin{align}\label{eq:representation_1}
 S_{\mu\nu}(\vec{q},t) =& \frac{1}{V} \int_V\!\diff \vec{x}\! \int_V\!\diff \vec{x}'  \,G(\vec{x},\vec{x}',t) \nonumber \\
&\times e^{-i(\vec{q}+ \vec{Q}_\mu )\cdot \vec{x} }   e^{i (\vec{q}+\vec{Q}_\nu ) \cdot \vec{x}' } ,
\end{align}
and reversely
\begin{align}\label{eq:representation_2}
G(\vec{x},\vec{x}',t) =& \frac{1}{V} \sum_{\mu,\nu \in \mathbb{Z}} \sum_{\vec{q}\in \text{BZ} } S_{\mu\nu}(\vec{q},t) \nonumber \\
&\times e^{i(\vec{q}+ \vec{Q}_\mu )\cdot \vec{x} }   e^{-i (\vec{q}+\vec{Q}_\nu ) \cdot \vec{x}' } .
\end{align}

\subsection{Currents and the continuity equation}
In this subsection we introduce the continuity equation for the particle density and define the associated currents. Their definition depends on the underlying dynamics. Although we have as main application a colloidal monolayer in mind, we formulate the theory here for Newtonian dynamics to keep the derivation of the subsequent mode-coupling theory  as simple as possible. 

From Hamilton's equations of motion we derive 
the continuity equation in real space for the particle density 
\begin{align}
\partial_t \rho(\vec{x},t) + \vec{\nabla} \cdot \vec{j}(\vec{x},t) = 0 ,
\end{align}
with the microscopic current 
\begin{align}
\vec{j}(\vec{x},t) := \sum_{n=1}^N \frac{\vec{p}_n(t)}{m} \delta[\vec{x} - \vec{x}_n(t) ].
\end{align}
In the spatial Fourier domain the current becomes
\begin{align}
\vec{j}(\vec{q}_\mu,t) = \vec{j}_\mu(\vec{q},t) := \sum_{n=1}^N \frac{\vec{p}_n(t)}{m} \exp[i (\vec{q}+\vec{Q}_\mu ) \cdot \vec{x}_n(t) ] .  
\end{align}
 The continuity equation  in terms of these variables then reads
 \begin{align}\label{eq:continuity}
 \partial_t \rho_\mu(\vec{q},t)  &= i (\vec{q} + \vec{Q}_\mu) \cdot \vec{j}_\mu(\vec{q},t) .
 \end{align}
The breaking of rotational symmetry due to the periodic modulation suggests splitting the wavevector $\vec{q}_\mu= \vec{q} + \vec{Q}_\mu$ into a component parallel to the modulation $\vec{q}_\mu^\parallel := (\vec{q}_\mu \cdot \vec{e}_z) \vec{e}_z$ and a perpendicular one $\vec{q}_\mu^\perp := \vec{q}_\mu - \vec{q}_\mu^\parallel$.  Note that $\vec{q}_\mu^\parallel = \vec{q}^\parallel + \vec{Q}_\mu$ and $\vec{q}_\mu^\perp = \vec{q}^\perp$ such that the direction of the associated unit vectors $\hat{\vec{q}}_\mu^\parallel = \hat{\vec{q}}^\parallel = \vec{q}^\parallel/q^\parallel = \vec{e}_z, \hat{\vec{q}}_\mu^\perp = \hat{\vec{q}}^\perp = \vec{q}^\perp/ q^\perp$ depends only on the direction $\vec{q} \in \text{BZ}$.  The currents therefore naturally split into 
\begin{subequations}
\begin{align}
j_\mu^\parallel(\vec{q},t) = \sum_{n=1}^N \frac{\hat{\vec{q}}^\parallel\cdot \vec{p}_n(t)}{m} \exp[i (\vec{q}+ \vec{Q}_\mu) \cdot \vec{x}_n(t) ], \\
j_\mu^\perp(\vec{q},t) =  \sum_{n=1}^N \frac{\hat{\vec{q}}^\perp \cdot \vec{p}_n(t)}{m}    \exp[i (\vec{q}+ \vec{Q}_\mu) \cdot \vec{x}_n(t) ]  .
\end{align} 
\end{subequations}
 We can unify the notation to
\begin{align}
j_\mu^\alpha(\vec{q},t) = \frac{1}{m} \sum_{n=1}^N & b^\alpha[\hat{\vec{q}}^\parallel\cdot\vec{p}_n(t), \hat{\vec{q}}^\perp\cdot\vec{p}_n(t)]    \nonumber  \\
& \times \exp[i (\vec{q}+ \vec{Q}_\mu) \cdot \vec{x}_n(t) ],
\end{align}
where the selector $b^\alpha(x,z) := x \delta_{\alpha,\perp}+ z \delta_{\alpha,\parallel}$ has been introduced. 
We shall use Greek letters $\alpha,\beta,..$ to indicate the relaxation channel and refer to these as \emph{channel indices}.   
 The continuity equation then reads in terms of these variables
 \begin{align}\label{eq:continuity}
 \partial_t \rho_\mu(\vec{q},t) &= i (q^\parallel + Q_\mu) j^\parallel_\mu(\vec{q},t) + i q^\perp j^\perp(\vec{q},t) \nonumber \\
 &=  i \sum_{\alpha=\parallel,\perp} (q+ Q_\mu)^\alpha j_\mu^\alpha(\vec{q},t) .
 \end{align}
 Here the superscript $\alpha=\parallel, \perp$ refers to the component parallel, resp. perpendicular to the modulation.

 We shall also need the correlation function of the currents at equal times 
 \begin{align}
 \mathcal{K}_{\mu\nu}^{\alpha\beta}(\vec{q}) =  \frac{1}{N} \langle {j}_\mu^\alpha(\vec{q},0)^* j_\nu^\beta(\vec{q},0) \rangle. 
 \end{align} 
 A simple calculation yields (see Appendix \ref{Sec:Static_current}) 
 \begin{align}
 \mathcal{K}_{\mu\nu}^{\alpha\beta}(\vec{q}) = v_{\text{th}}^2  \delta^{\alpha\beta} \frac{n_{\nu-\mu}}{n_{0}} ,
 \end{align}
with the thermal velocity $v_{\text{th}} = \sqrt{k_B T/m}$ and $k_B  T $ the thermal energy. 
The inverse matrix is then found using the convolution relation, Eq.~\eqref{eq:convolution_theorem}, 
 \begin{align}
 [\bm{\mathcal{K}}^{-1}(\vec{q}) ]^{\alpha\beta}_{\mu\nu} = \frac{1}{v_{\text{th}}^2} n_{0} \delta^{\alpha\beta} v_{\nu-\mu} .
 \end{align}
 
\section{Mori-Zwanzig equations of motion}\label{Sec:Mori_Zwanzig}
We use the Mori-Zwanzig projection operator formalism~\cite{Hansen:Theory_of_Simple_Liquids,Goetze:Complex_Dynamics} to derive formally exact equations of motion for 
the generalized intermediate scattering functions. In Hamiltonian mechanics the time evolution of an observable  $A(t) \equiv A(\vec{x}(t), \vec{p}(t)) $ is driven by the canonical equations 
of motion $\diff A(t)/\diff t = \{ A(t), \mathcal{H}\}$ where $\{\, . , .\}$ denotes the Poisson bracket. 
It is convenient to introduce the  Liouville operator $\mathcal{L}$ via $i \mathcal{L} = \{ \, .\, , \mathcal{H} \}$. 
The formal solution then is $A(t) = \exp(i \mathcal{L} t ) A$ where we adopt the convention that if no time argument is provided, the observable is evaluated at time $t=0$.  

The set of fluctuating observables $\delta A := A- \langle A \rangle$ naturally acquires the structure of a  Hilbert space  upon defining the Kubo scalar product $\langle A | B \rangle := \langle \delta A^* \delta B \rangle$.  One readily shows that the Liouville operator is Hermitian with respect to the Kubo scalar product $\langle A |\mathcal{L} B \rangle = \langle \mathcal{L} A | B \rangle$. Time-dependent correlation functions can therefore be represented as matrix elements $\langle \delta A(t)^* \delta B \rangle = \langle A | \mathcal{R}(t) | B \rangle$ of the backwards-time evolution operator $\mathcal{R}(t) =\exp(-i \mathcal{L} t)$. 

The starting point of the Zwanzig-Mori formalism relies on the operator identity (see e.g.~\cite{Lang:PRE_86:2012})
\begin{align}\label{eq:operator_identity}
\partial_t \mathcal{P} &\mathcal{R}(t) \mathcal{P} + i \mathcal{P} \mathcal{L} \mathcal{P} \mathcal{R}(t) \mathcal{P}  \nonumber \\
&+ \int_0^t \diff t' \mathcal{P} \mathcal{L} \mathcal{Q} e^{-i \mathcal{Q}\mathcal{L}\mathcal{Q} (t-t')} \mathcal{Q}\mathcal{L}\mathcal{P} \mathcal{R}(t') \mathcal{P} =0 ,
\end{align}
valid for any orthogonal projector
 $\mathcal{P} = \mathcal{P}^\dagger = \mathcal{P}^2$  and $\mathcal{Q}= 1-\mathcal{P}$ its orthogonal complement. 
 
The derivation of  the equations of motion (e.o.m.\@) largely parallels the one for the slit geometry~\cite{Lang:PRE_86:2012}. 
 First we choose the projection operator on the densities
 \begin{align}
 \mathcal{P}_\rho := \frac{1}{N} \sum_{\mu,\nu\in \mathbb{Z} } 
\sum_{ \vec{q}\in \text{BZ} } 
|\rho_\mu(\vec{q}) \rangle 
[\mathbf{S}^{-1}(\vec{q}) ]_{\mu\nu} \langle \rho_\nu(\vec{q}) | ,
 \end{align}
 where we introduced the matrix notation $[\mathbf{S}(\vec{q})]_{\mu\nu} = S_{\mu\nu}(\vec{q})$. One readily checks that it fulfills the properties of an orthogonal projector. Then sandwiching the operator identity between bras and kets consisting of density modes 
one finds the e.o.m.\@
\begin{align}\label{eq:ISF1}
\dot{S}_{\mu\nu}(\vec{q},t)  + &\sum_{\kappa,\lambda\in \mathbb{Z} } \int_0^t K_{\mu\kappa}(\vec{q},t-t') \nonumber \\
& \times [\mathbf{S}^{-1}(\vec{q})]_{\kappa\lambda} S_{\lambda \nu}(\vec{q},t') \diff t' = 0 .
\end{align} 
To simplify notation, we switch to a  matrix notation   and suppress the dependence on the wave vector $\vec{q}\in \text{BZ}$ in the equation as it is unchanged throughout the equation 
\begin{align}\label{eq:ISF2}
\dot{\mathbf{S}}(t) + \int_0^t \mathbf{K}(t-t') \mathbf{S}^{-1} \mathbf{S}(t') \diff t' = 0.
\end{align}
The microscopic expression for the memory kernel $[\mathbf{K}(\vec{q},t) ]_{\mu\nu} = K_{\mu\nu}(\vec{q},t)$ is provided by
\begin{align}
K_{\mu\nu}(\vec{q},t) = \frac{1}{N}\langle \mathcal{L} \rho_\mu(\vec{q}) | e^{-i \mathcal{Q}_\rho \mathcal{L} \mathcal{Q}_\rho t } | \mathcal{L} \rho_\nu(\vec{q}) \rangle.
\end{align} 
In the jargon of the Mori-Zwanzig projection-operator formalism the operator $\exp( -i \mathcal{Q}_\rho \mathcal{L} \mathcal{Q}_\rho t)$ drives  the reduced (backwards) time evolution with respect to the projector $\mathcal{P}_\rho$. 

Inserting the particle-conservation law, Eq.~\eqref{eq:continuity}, we find that the current kernel naturally splits
\begin{align}\label{eq:current_kernel}
K_{\mu\nu}(\vec{q},t) = \sum_{\alpha,\beta= \parallel,\perp} (q+Q_\mu)^\alpha \mathcal{K}_{\mu\nu}^{\alpha\beta}(\vec{q},t) (q+ Q_\nu)^\beta,
\end{align} 
with 
\begin{align}
\mathcal{K}_{\mu\nu}^{\alpha\beta}(\vec{q},t) = \frac{1}{N} \langle j_\mu^\alpha(\vec{q}) | e^{-i \mathcal{Q}_\rho \mathcal{L} \mathcal{Q}_\rho t } | j_\nu^\beta(\vec{q}) \rangle . 
\end{align} 
Note that the initial value reduces to the static current correlation function $\mathcal{K}_{\mu\nu}^{\alpha\beta}(\vec{q},t=0) = \mathcal{K}_{\mu\nu}^{\alpha\beta}(\vec{q})$.

Next we derive an exact equation of motion for this current kernel. We introduce the projector on the currents
\begin{align}
\mathcal{P}_j = N^{-1} \sum_{\alpha,\beta=\parallel,\perp}\sum_{\mu,\nu\in \mathbb{Z}} \sum_{\vec{q}\in\text{BZ}} |j_\mu^\alpha(\vec{q}) \rangle [\boldmath{\mathcal{K}}^{-1}(\vec{q}) ]^{\alpha\beta}_{\mu\nu} \langle j_\nu^\beta(\vec{q}) |  .
\end{align}
Note that the projection operators commute $\mathcal{P} := \mathcal{P}_\rho\mathcal{P}_j = \mathcal{P}_j \mathcal{P}_\rho$. Sandwiching the operator identity, Eq.~\eqref{eq:operator_identity} between current modes, we find another exact equation of motion
\begin{align}
\dot{\mathcal{K}}^{\alpha\beta}_{\mu\nu}(\vec{q},t)    +  & \sum_{\gamma,\delta=\parallel,\perp}\sum_{\kappa,\lambda\in \mathbb{Z}}\int_0^t \mathfrak{M}^{\alpha\gamma}_{\mu\kappa}(\vec{q},t-t')  \nonumber \\
& \times  [\bm{\mathcal{K}}^{-1}(\vec{q})]^{\gamma\delta}_{\kappa\lambda} \mathcal{K}_{\lambda\nu}^{\delta\beta}(\vec{q},t') \diff t' =0 .
\end{align} 
Here the force kernel is given by
\begin{align}
\mathfrak{M}^{\alpha\beta}_{\mu\nu}(\vec{q},t) = \frac{1}{N}\langle \mathcal{Q} \mathcal{L} j^\alpha_\mu(\vec{q}) | e^{-i \mathcal{Q}\mathcal{L} \mathcal{Q} t} | \mathcal{Q} \mathcal{L} j^\beta_\nu(\vec{q}) \rangle ,
\end{align}
with the projection operator $\mathcal{Q} := \mathcal{Q}_j \mathcal{Q}_\rho$.  Using a matrix notation with superindices consisting of the pair of a channel index and a mode index $(\alpha,\mu)$ we can write the e.o.m.\@ as
\begin{align}\label{eq:eom_zwanzig_currents}
\dot{\bm{\mathcal{K}}}(t) + \int_0^t \bm{\mathfrak{M}}(t-t') \bm{\mathcal{K}}^{-1} \bm{\mathcal{K}}(t') \diff t' =0 .
\end{align} 

It is also convenient to formulate the e.o.m\@ in the Laplace domain since  then the time convolutions become local in the frequency domain. We define Laplace transforms using the  convention
\begin{align}
\hat{S}_{\mu\nu}(q,z) = i \int_0^\infty S_{\mu\nu}(q,t) e^{i z t} \diff t, \qquad \Imag[z] > 0,
\end{align} 
and similarly for all other quantities.
The e.o.m.\@ assume the following form
\begin{align}\label{eq:eom_ISF_Laplace}
\hat{\mathbf{S}}(z) = -[z \mathbf{S}^{-1} + \mathbf{S}^{-1} \hat{\mathbf{K}}(z) \mathbf{S}^{-1} ]^{-1} ,
\end{align}
and
\begin{align}\label{eq:eom_current_Laplace}
\hat{\bm{\mathcal{K}}}(z) = - [z \bm{\mathcal{K}}^{-1} + \bm{\mathcal{K}}^{-1} \hat{\bm{\mathfrak{M}}}(z) \bm{\mathcal{K}}^{-1}]^{-1} . 
\end{align}

The equations of motion as well as the formally exact equations for the memory kernel are virtually identical to the ones for a slit geometry except that the wave vector $\vec{q}\in \text{BZ}$ in general cannot be reduced to a scalar by suitable rotations. Employing the symmetries of a modulated liquid allows only  restricting $\vec{q}$ to a two-dimensional cut of the degenerate Brillouin zone consisting of a single perpendicular directions and the parallel direction $ -\pi / a <q_z \leq \pi/a$.   

\section{Mode-coupling theory}\label{Sec:MCT}

The e.o.m.\@ derived in the previous section are exact, however, in a certain sense also empty since they merely reformulate the original problem for the generalized intermediate scattering functions in terms of unknown memory kernels. The strategy of the mode-coupling approach~\cite{Goetze:Complex_Dynamics} is to close the equations of motion by providing an approximate relation between the unknown memory kernel and the intermediate scattering functions itself. To this end, the fluctuating force variables are represented as a linear superposition of density pair modes, concomitantly the reduced dynamics of the pair modes are factorized into a product of pair modes evolving in time by the original dynamics. As a result, 
the mode-coupling procedure leads to an approximation of the force kernel as a bilinear functional of the generalized intermediate scattering functions. 

Following literally the procedure of Ref.~\cite{Lang:PRE_86:2012}, the mode-coupling approximation leads to
\begin{align}
\mathfrak{M}^{\alpha\beta}_{\mu\nu}(\vec{q},t) \approx & \frac{1}{2 N^3} 
\sum_{\vec{q}_1,\vec{q}_2 \in \text{BZ}} 
\sum_{\substack{\mu_1,\mu_2 \\ \nu_1,\nu_2}\in \mathbb{Z}} 
\mathcal{X}^\alpha_{\mu,\mu_1\mu_2}(\vec{q},\vec{q}_1\vec{q}_2)  \nonumber \\
&\times S_{\mu_1\nu_1}(\vec{q}_1,t) S_{\mu_2\nu_2}(\vec{q}_2,t) \mathcal{X}^\beta_{\nu,\nu_1\nu_2}(\vec{q},\vec{q}_1\vec{q}_2)^* ,
\end{align} 
where the complex-valued vertices arise from the overlap of the fluctuating forces with the density pair modes
\begin{align}
\mathcal{X}^\alpha_{\mu,\mu_1\mu_2}(\vec{q},\vec{q}_1\vec{q}_2) =& \sum_{\mu_1',\mu_2'\in \mathbb{Z}} \langle [\mathcal{Q} \mathcal{L} j_\mu^\alpha(\vec{q})]^* \delta \rho_{\mu_1'}(\vec{q}_1) \delta\rho_{\mu_2'}(\vec{q}_2) \rangle\nonumber \\
& \times [\mathbf{S}^{-1}(\vec{q}_1) ]_{\mu_1'\mu_1} [\mathbf{S}^{-1}(\vec{q}_2)]_{\mu_2'\mu_2}.
\end{align}
The overlap can be expressed in terms of structural quantities only \begin{align}\label{eq:overlap}
&\langle [\mathcal{Q} \mathcal{L} j_\mu^\alpha(\vec{q})]^* \delta \rho_{\mu_1'}(\vec{q}_1) \delta\rho_{\mu_2'}(\vec{q}_2) \rangle \nonumber\\ 
=&N v_\text{th}^2  \sum_{\mu^\sharp=0,\pm 1} \delta_{\vec{q}-\vec{Q}_{\mu^\sharp},\vec{q}_{1}+\vec{q}_{2}}\big\{
(q_1+ Q_{\mu_1})^\alpha
S_{\mu-\mu_{1}+\mu^\sharp,\mu_{2}}(\vec{q}_{2})\nonumber \\
&+(1\leftrightarrow2)\nonumber -\sum_{\mu',\mu'' \in \mathbb{Z}} \frac{n_{\mu'-\mu}}{n_{0}}
(q+Q_{\mu'})^\alpha
 [\mathbf{S}^{-1}(\vec{q})]_{\mu'\mu''} \nonumber \\
& \times S_{\mu'',\mu_{1}\mu_{2}}(\vec{q},\vec{q}_{1}\vec{q}_{2})\big\},
\end{align}
see Appendix \ref{sec:scalarproduct_a}. Here in addition to the generalized static structure factors also the triple correlation function
\begin{align}
S_{\mu,\mu_1\mu_2}(\vec{q},\vec{q}_1\vec{q}_2) = \frac{1}{N} \langle \delta \rho_\mu(\vec{q})^* \delta \rho_{\mu_1}(\vec{q}_1) \delta \rho_{\mu_2}(\vec{q}_2) \rangle , 
\end{align}
enters the expression. The triple correlation function is hard to obtain in praxis and to make progress additional approximations are invoked. Here we rely on the convolution approximation for  inhomogeneous liquids as introduced by Rajan \emph{et al}~\cite{Rajan:JMath19:1978} generalizing the standard bulk convolution approximation~\cite{Hansen:Theory_of_Simple_Liquids}. The application to the case of periodically modulated liquid reflecting the discrete translational symmetry is elaborated in Appendix~\ref{sec:convolution2}. Using this result the expression for the vertex drastically simplifies to 
\begin{align}
&  \mathcal{X}^{\alpha}_{\mu,\mu_{1}\mu_{2}} 
  (\vec{q},\vec{q}_{1}\vec{q}_{2})  \approx -N n_{0} v_\text{th}^2 \sum_{\mu^\sharp = 0, \pm 1} 
  \delta_{\vec{q}-\vec{Q}_{\mu^\sharp} ,\vec{q}_{1}+\vec{q}_{2}}  \nonumber \\
  \times & \left[ (q_1+ Q_{\mu+\mu^\sharp}-Q_{\mu_2})^\alpha
    c_{\mu+ \mu^\sharp-\mu_2,\mu_1}(\vec{q}_1) 
+( 1 \leftrightarrow 2 ) \right], 
\end{align}
see Appendix~\ref{sec:vertex_approximation} for a derivation. The vertex is expressed in terms of the Fourier representation $c_{\mu\nu}(\vec{q})$  of the direct correlation function defined via the inhomogeneous Ornstein-Zernike relation
~\cite{Hansen:Theory_of_Simple_Liquids}, see  Appendix~\ref{sec:Ornstein_Zernike} for details. In the Fourier domain for a modulated liquid it is related to the matrix of the static structure factor via 
\begin{align}\label{eq:Ornstein-Zernike}
[\mathbf{S}^{-1}(\vec{q}) ]_{\mu\nu} = n_{0} v_{ \nu-\mu} - n_{0} c_{\mu\nu}(\vec{q}).
\end{align}

The force kernel enters the Mori-Zwanzig equation of motion for the current kernel in the Laplace domain, Eq.~\eqref{eq:eom_current_Laplace},  only in the combination  $ \bm{\mathcal{K}}^{-1}(\vec{q})\hat{\bm{\mathfrak{M}}}(\vec{q},z) \bm{\mathcal{K}}^{-1}(\vec{q})$ which suggests introducing the effective force kernel 
\begin{align}\label{eq:effective}
\mathcal{M}^{\alpha\beta}_{\mu\nu}(\vec{q},t) &:=[\bm{\mathcal {K}}^{-1}(\vec{q})\mathbf {\mathfrak{M}}(\vec{q},t)\bm{\mathcal {K}}^{-1}(\vec{q})]^{\alpha\beta}_{\mu\nu} \nonumber  \\ 
&\approx\mathcal{F}_{\mu\nu}^{\alpha\beta}[\mathbf{S}(t),\mathbf{S}(t);\vec{q}] \nonumber\\
&=\frac{1}{2N} 
\sum_{\vec{q}_1\vec{q}_2 \in \text{BZ}} 
\sum_{\substack{\mu_1\mu_2 \\ \nu_1\nu_2}\in \mathbb{Z}} 
\mathcal{Y}^{\alpha}_{\mu;\mu_{1}\mu_{2}}(\vec{q},\vec{q}_{1}\vec{q}_{2})\nonumber\\
&\times S_{\mu_{1}\nu_{1}}(\vec{q}_{1},t) S_{\mu_{2}\nu_{2}}(\vec{q}_{2},t) \mathcal{Y}^{\beta}_{\nu;\nu_{1}\nu_{2}}(\vec{q},\vec{q}_{1}\vec{q}_{2})^*, 
\end{align}
with new vertices
\begin{align}\label{eq:Y_vertices}
&\mathcal{Y}^{\alpha}_{\mu,\mu_{1}\mu_{2}}(\vec{q},\vec{q}_{1}\vec{q}_{2})\nonumber \\
=& n_{0}^2  \sum_{\mu^\sharp = 0, \pm 1} 
  \delta_{\vec{q}-\vec{Q}_{\mu^\sharp} ,\vec{q}_{1}+\vec{q}_{2}}  \sum_{\kappa\in \mathbb{Z} } v^*_{\mu-\kappa}
  \nonumber \\
  &\times  \left[ (q_1+ Q_{\kappa+\mu^\sharp}-Q_{\mu_2})^\alpha
    c_{\kappa+ \mu^\sharp-\mu_2,\mu_1}(\vec{q}_1) 
+( 1 \leftrightarrow 2 ) \right]
.
\end{align}
The explicit expression of the mode-coupling functional and the vertices, Eqs.~\eqref{eq:effective},\eqref{eq:Y_vertices}, together with the equations of motion of Sec.~\ref{Sec:Mori_Zwanzig} is the main result of this work. 
The structure of  the vertices is reminiscent of the cases of a confined liquid~\cite{Lang:PRE_86:2012} or a liquid of comprising of molecules~\cite{Scheidsteger:PRE56:1997,Franosch:PRE56:1997}. 
As additional feature of the modulation we find that only modes can couple  satisfying the selection rule $\vec{q}-\vec{q}_1-\vec{q}_2 \in \Lambda^*$ which reflects the conservation of crystal momentum. 

The notation for the mode-coupling functional $\mathcal{F}_{\mu\nu}^{\alpha\beta}[\mathbf{S}(t),\mathbf{S}(t);\vec{q}] $ in Eq.~\eqref{eq:effective} highlights that the functional is bilinear in the matrices of the intermediate scattering functions, which is a general feature of the mode-coupling approach~\cite{Goetze:Complex_Dynamics}. The structure of the Mori-Zwanzig equations of motion, Eqs.~\eqref{eq:eom_ISF_Laplace} and \eqref{eq:eom_current_Laplace},
together with the MCT closure, 
Eq.~\eqref{eq:effective},  falls into the class of general MCT equations with multiple decay 
channels~\cite{Lang:JSTATMECH_12:2013} for which various mathematical properties hold. In particular, for given parameters, i.e. the density profile $n_\mu$ and static structure factors $S_{\mu\nu}(\vec{q})$ a unique solution of the MCT equations exists and the solution respects the mathematical properties of a correlation function. 

\section{Nonergodicity parameters}\label{Sec:Nonergodicity} 

Here  we show that the theory allows for a glass transition as encoded in a nonergodicity transition. Since our theory belongs to the class of MCT equations with multiple decay channels, this section merely summarize known results~\cite{Lang:PRE_86:2012,Lang:JSTATMECH_12:2013} and serves to fix the notation but also as a reference for future work. 

The long-time limits of the intermediate scattering functions are denoted by 
\begin{align}
F_{\mu\nu}(\vec{q}) := \lim_{t\to\infty} S_{\mu\nu}(\vec{q},t) ,
\end{align}
and referred to as nonergodicity parameters.
This limit is known to exist for pure relaxational dynamics~\cite{Goetze:JMath195:1995, Franosch:JStatPhys109:2002, Jung:JSTATMECH:2020} and also for a broad class of Newtonian dynamics \cite{Franosch:JMathA:2014}. Liquid states are defined if $F_{\mu\nu}(\vec{q}) \equiv 0$ for all wave vectors $\vec{q}$ and mode indices $\mu,\nu$, and are referred to as ergodic~\cite{Goetze:Complex_Dynamics}. In contrast, glassy states are nonergodic $F_{\mu\nu}(\vec{q}) \neq 0$. A convenient representation is obtained as the low-frequency limit in the Laplace domain
\begin{align}
F_{\mu\nu}(\vec{q}) = - \lim_{z\to 0} z \hat{S}_{\mu\nu}(\vec{q},z)  .
\end{align}
where the limit $z\to 0$ is always understood to be  performed in a sector in the upper half plane $ \delta < \text{arg }(z) < \pi -\delta, \delta>0$. Since the matrix $\mathbf{S}(\vec{q},t)$ is a matrix-valued autocorrelation function, the associated matrix of nonergodicity parameters $\mathbf{F}(\vec{q}) $ is positive-semidefinite for each wave vector $\vec{q}$.

By the Mori-Zwanzig equations of motion, a nonergodic state can happen only if the long-time limit of the force kernels
\begin{align}\label{eq:N_limit} 
\bm{\mathcal{N}}(\vec{q}) := \lim_{t\to \infty} \bm{\mathcal{M}}(\vec{q},t) = \bm{\mathcal{F}}[\mathbf{F},\mathbf{F};\vec{q}]  , 
\end{align}
becomes nonergodic simultaneously. The e.o.m.\@ for the currents, Eq.~\eqref{eq:eom_current_Laplace}, then shows that the low-frequency behavior of the current kernel in the Laplace domain is provided by $\hat{K}_{\mu\nu}(\vec{q},z) = z G_{\mu\nu}(\vec{q}) + o(z)$ for $z\to 0$ with
\begin{align}\label{eq:G_limit}
G_{\mu\nu}(\vec{q}) := \sum_{\alpha,\beta = \parallel, \perp} 
 (q+Q_\mu)^\alpha [ \bm{\mathcal{N}}^{-1}(\vec{q})]^{\alpha\beta}_{\mu\nu}  (q+ Q_\nu)^\beta.
\end{align}
Similarly, the e.o.m., Eq,~\eqref{eq:eom_ISF_Laplace}, reveals that the nonergodicity  parameters are connected to the low-frequency behavior of the currents via
\begin{align}\label{eq:F_to_G}
\mathbf{F}(\vec{q}) = \mathbf{S}(\vec{q}) - [ \mathbf{S}^{-1}(\vec{q})+ \mathbf{G}^{-1}(\vec{q}) ]^{-1}. 
\end{align} 
Equations~\eqref{eq:N_limit},\eqref{eq:G_limit},\eqref{eq:F_to_G} have to be solved self-consistently. A vanishing nonergodicity matrix $\mathbf{F}(\vec{q})\equiv 0$ is always a solution (in which case we formally associate the infinite matrix $\mathbf{G}(\vec{q})$ with $\mathbf{G}^{-1}(\vec{q})=0$), however nontrivial solutions may occur, provided the coupling encoded in the mode-coupling functional $\bm{\mathcal{F}}[\mathbf{F},\mathbf{F}; \vec{q}]$ is sufficiently strong. The unique solution as obtained from the long-time limit of the solution of the e.o.m.\@ is obtained by a convergent iteration scheme~\cite{Lang:PRE_86:2012,Lang:JSTATMECH_12:2013} without resorting to the full time-dependent solution.

\section{Comparison to bulk liquids}\label{Sec:bulk_MCT}
In this section, we demonstrate that in the absence of an external modulation the theory simplifies to the conventional MCT equations governing the glass transition in bulk systems. 

The complete translational and rotational symmetry is restored in bulk liquids leading to a uniform equilibrium density, $n(\vec{x})=n_{0}=1/v= \text{const.}$, which implies for the Fourier coefficients of the density and volume
\begin{subequations}
\begin{align}
 n_{\mu} &= n_{0} \delta_{\mu 0}, \\
 v_{\mu} &= v \delta_{\mu 0}.
\end{align}
\end{subequations}
Similarly, the intermediate scattering function becomes diagonal with respect to the mode indices
\begin{equation}
 S_{\mu\nu}(\vec{q},t) = \delta_{\mu\nu} S(k,t),
\end{equation}
where $S(k,t)$ refers to the conventional intermediate scattering function of a bulk system and we abbreviate in this section $\vec{k} = \vec{q}+\vec{Q}_{\mu}$. By isotropy it depends only on the magnitude $k=|\vec{k}|$ of the wave vector. 
Specializing to time $t=0$ implies the same relation for the generalized structure factor. Then the Ornstein-Zernike relation, Eq.~\eqref{eq:Ornstein-Zernike}, reveals
\begin{equation}\label{eq:cbulk}
 c_{\mu\nu}(\vec{q}) = \delta_{\mu\nu} c(k),
\end{equation}
where $c(k) \in \mathbb{R}$ is the ordinary bulk direct correlation function. 

Similar considerations hold for the tensorial correlation functions. 

For bulk liquids, isotropy holds which implies for the effective force kernel  
\begin{align}\label{eq:tensorial}
\mathcal{M}^{\alpha\beta}_{\mu\nu}(\vec{q},t) = \delta_{\mu\nu} \left[ 
\hat{k}^\alpha \hat{k}^\beta \mathcal{M}^L (k,t)
+(\delta^{\alpha\beta} - \hat{k}^\alpha \hat{k}^\beta) \mathcal{M}^T (k,t) \right]  ,
\end{align}
where $\mathcal{M}^L(k,t)$ and $\mathcal{M}^T(k,t)$ denote the longitudinal and transversal components and $\hat{k}^\alpha = k^\alpha/k$. Similar relations hold for the current kernel $\mathcal{K}^{\alpha\beta}_{\mu\nu}(\vec{q},t)$. Note that by Eq.~\eqref{eq:current_kernel} only the longitudinal part $\mathcal{K}^L(k,t)$ contributes to $K_{\mu\nu}(\vec{q},t) = \delta_{\mu\nu} K(k,t)$ with $K(k,t) =k^2 \mathcal{K}^L(k,t)$.

In the  expression for the vertex, Eq.~\eqref{eq:Y_vertices}, only the term $\kappa = \mu$ contributes due to the constant volume per particle. Furthermore $\mu_1 = \mu+ \mu^\sharp-\mu_2$ is enforced by the diagonality of the direct correlation function, which also implies $(q_1 + Q_{\mu+\mu^\sharp} - Q_{\mu_2})^\alpha = k_1^\alpha$. Last, with $\sum_{\mu^\sharp= 0,\pm 1} \delta_{\vec{q}-\vec{Q}_{\mu^\sharp}, \vec{q}_1+\vec{q}_2} = \delta_{\vec{k}, \vec{k}_1+ \vec{k}_2}$ and upon collecting results we find
\begin{align}
\mathcal{Y}^{\alpha}_{\mu,\mu_{1}\mu_{2}}(\vec{q},\vec{q}_{1}\vec{q}_{2}) 
&= n_{0} \delta_{\vec{k},\vec{k}_1+ \vec{k}_2} [ k_1^\alpha c(k_1)+(1\leftrightarrow 2)] 
\nonumber
\\   &\equiv \mathcal{Y}^{\alpha}(\vec{k},\vec{k}_{1}\vec{k}_{2}).
\end{align}
Note that the Kronecker delta reflects momentum conservation for fully restored translational invariance. Furthermore $\mathcal{Y}^{\alpha}(\vec{k},\vec{k}_{1}\vec{k}_{2})$ transforms as a vector under rotations.

 For the memory kernel, Eq.~\eqref{eq:effective}, momentum conservation implies that it is only nonvanishing for $\mu = \nu$. 
Observing that $\sum_{\vec{q}_1\in \text{BZ}} \sum_{\mu_1\in \mathbb{Z}} \ldots = \sum_{\vec{k}\in \lambda^*} \ldots$, we find for the force kernel in 
 MCT approximation
\begin{align}
&\mathcal{M}^{\alpha\beta}_{\mu\nu}(\vec{q},t)= \delta_{\mu\nu} \frac{1}{2N} \sum_{\vec{k}_1,\vec{k}_2 \in \lambda^*}  \mathcal{Y}^{\alpha}(\vec{k},\vec{k}_{1}\vec{k}_{2})\nonumber\\
&\times S(k_{1},t)  S(k_{2},t) \mathcal{Y}^{\beta}(\vec{k},\vec{k}_{1}\vec{k}_{2})^*.
\end{align}
In particular,  the MCT approximation does not spoil the tensorial character of the effective force kernel, Eq.~\eqref{eq:tensorial}. 
We  find for the longitudinal part 
\begin{align}
\mathcal{M}^L(k,t) =& \frac{n_{0}^2}{2N} \sum_{\vec{k}_1, \vec{k}_2 \in \lambda^*} \delta_{\vec{k},\vec{k}_1+\vec{k}_2} [ \hat{\vec{k}} \cdot\vec{k}_1 c(k_1) + (1 \leftrightarrow 2) ]^2 \nonumber \\
& \times S(k_1,t) S(k_2,t) .
\end{align}
The previous expression coincides with the conventional mode-coupling approximation for bulk liquids~\cite{Goetze:Complex_Dynamics, Hansen:Theory_of_Simple_Liquids}.

Next we show that the e.o.m, Eqs.~\eqref{eq:ISF2},
\eqref{eq:eom_zwanzig_currents}, reduce to  the well-known equations for bulk liquid. First, for the generalized intermediate scattering function we find directly from Eq.~\eqref{eq:ISF2}
\begin{align}
\dot{S}(k,t) + \int_0^t K(k,t-t') S(k)^{-1} S(k,t') \diff t' = 0,
\end{align}
with formal solution in the Laplace domain
\begin{align}\label{eq:bulk_eom1}
\hat{S}(k,z) = - [ z S(k)^{-1} + S(k)^{-1} \hat{K}(k,z) S(k)^{-1} ]^{-1} .  
\end{align}
Then we decompose the e.o.m.\@ for the current kernel, Eqs.~\eqref{eq:eom_zwanzig_currents}, in terms of the effective force kernel
\begin{align}
\dot{\bm{\mathcal{K}}}(\vec{q},t) + \int_0^t  \bm{\mathcal{K}}(\vec{q}) {\mathcal{M}}(\vec{q},t-t')  \bm{\mathcal{K}}(\vec{q},t') \diff t' =0, 
\end{align}
into its longitudinal and transverse part. Since the projection operators $\hat{k}^\alpha \hat{k}^\beta$ and $\delta^{\alpha\beta} - \hat{k}^\alpha \hat{k}^\beta$ onto the longitudinal and transversal component 
are orthogonal, and for an isotropic liquid $\mathcal{K}^{\alpha\beta}_{\mu\nu}(\vec{q}) = v_{\text{th}}^2 \delta^{\alpha\beta} \delta_{\mu\nu}$, we find
\begin{align}  
\dot{\mathcal{K}}^L(k,t) +  v_{\text{th}}^2 \int_0^t \mathcal{M}^L(k,t-t') \mathcal{K}^L(k,t') \diff t' = 0,
\end{align}
and similarly for the transversal parts. By Laplace transform, the solution is formally expressed as
\begin{align}\label{eq:bulk_eom2}
\hat{\mathcal{K}}(k,z) = - [z /v_{\text{th}}^2 + \hat{\mathcal{M}}^L(k,z) ]^{-1} . 
\end{align} 
Observing $\hat{K}(k,z) = k^2 \hat{\mathcal{K}}(k,z)$, Eqs.~\eqref{eq:bulk_eom1},\eqref{eq:bulk_eom2}, can be combined to the conventional 
form~\cite{Goetze:Complex_Dynamics, Hansen:Theory_of_Simple_Liquids}

\begin{align}
\hat{S}(k,z)/S(k)= \frac{-1}{   z - \Omega(k)^2 / [ z +  \Omega(k)^2 \hat{m}(k,z) ] }  ,
\end{align}
with the characteristic frequency  $\Omega(k)^2 =k^2  v_{\text{th}}^2 /S(k)$ and 
the memory kernel $\hat{m}(k,z) = \hat{\mathcal{M}}^L(k,z) S(k)/k^2$.  

In summary, we find that the mode-coupling theory for modulated liquids reduces to the well-known MCT for bulk liquids. In particular, our calculation demonstrates the covariance of the theory. For a bulk liquid the introduction of a lattice  should have no impact on the dynamics of the system, in particular, the orientation of the lattice and the lattice constant  should be irrelevant. Since the Mori-Zwanzig formalism is merely an exact reformulation of the problem, it is anticipated that this covariance is reflected in the equations of motion. The nontrivial observation is that also the mode-coupling approximation for the force kernel preserves the covariance. A similar covariance property for mode-coupling theory has been demonstrated for the case molecular liquids~\cite{Schilling:PRE:65:2002} where the choice of a reference point on the molecules should not affect the dynamics.

\section{Summary and Conclusions}\label{Sec:Summary}
We have derived a mode-coupling theory for the dynamics of a liquid exposed to an external one-dimensional  periodic potential. The liquid becomes inhomogeneous, in particular, a density modulation of the same period results. To account for the breaking of translational symmetry we have introduced appropriately generalized intermediate scattering functions reflecting the discrete translational symmetry, as familiar from the theory of crystalline solids.  The loss of rotational symmetry suggests splitting the currents into components parallel and perpendicular to the modulation giving rise to Mori-Zwanzig equations of motion of a more general structure than for bulk liquids. 
The force kernels are then approximated in terms of a mode-coupling approach which is known to  yield a successful   description of the structural relaxation in bulk systems but also for molecules and confined systems.

As a new feature of the discrete symmetry, momentum is not strictly conserved but may involve a reciprocal lattice vector. These \emph{Um\-klapp} process are properly reflected in the construction of the mode-coupling functional which corroborates that the mode-coupling approach respects all underlying symmetries of the system. Similar observations hold for the conservation of momentum in (quasi-)confined systems~\cite{Lang:PRE_86:2012, Schrack:JSTATMECH:2020} or angular momentum for molecular liquids~\cite{Franosch:PRE56:1997, Scheidsteger:PRE56:1997}. 

Whereas for molecular liquids the splitting of the currents was mandatory if the structural relaxation dynamics is assumed to be independent of inertial parameters such as masses and  moments of inertia~\cite{Schilling:PRE:65:2002},  
the splitting of the currents in modulated or (quasi-)confined liquids is suggested by physical intuition. As a benefit the static current-density matrix  
can be inverted without introducing additional structural information.  
We anticipate that the MCT functional for a modulated colloidal liquid will be identical to the one derived in our theory. The derivation based on Brownian dynamics will be addressed in future work. 

The present theory for modulated liquids shares all the mathematical properties derived for mode-coupling theories for multiple decay channels and correspondingly all mathematical properties can be directly inferred.  In particular, the existence of unique solutions can be demonstrated and moreover these solutions belong to the class of admissible functions for general stochastic processes. 

Formal equations for a mode-coupling theory for general inhomogeneous environments have been derived by Biroli \emph{et al} \cite{Biroli:PRL_97:2006} and recently also for generalized mode-coupling theory~ \cite{Laudicina:PE106:2022}. The main purpose there was to consider changes of the intermediate scattering functions in response to an infinitesimal external potential as encoded in higher-order density correlation functions in the absence of such a perturbation. These higher-order susceptibilities have been used to unravel divergent correlations and growing length scales as the glass transition is approached. By linear response, it is sufficient to apply a periodic modulation to derive these susceptibilities and correspondingly our theory should reproduce the results in the limiting case of small amplitude modulations.  

A one-dimensional liquid exposed to a periodic potential has been introduced by  Nandi \emph{et al }\cite{Nandi:2011} relying on a simplified mode-coupling approximation in order to discuss in a toy model the effects of confining hard walls. Our approach differs in several respects from their approach, in particular, we take into account all density modes as encoded in the matrix of the intermediate scattering functions  necessary to reconstruct the real-space density-density correlation functions. Correspondingly, our theory properly accounts for \emph{Um\-klapp} processes inherent to the residual discrete translational symmetry. 

We have presented the theory for the case of a one-dimensional potential modulation, yet the generalization to arbitrary periodic potentials is straightforward at the price of few notational changes. For example, the Brillouin zone will no longer be a stripe and the reciprocal lattice will have more than one dimension. The currents need to be split into several components compared to the one-dimensional modulation. While the experimental realization of such arbitrary periodic potentials appears  feasible,  it will be more and more difficult to preempt the crystallization as the modulation promotes crystallization,  a phenomenon also called 'light-induced freezing'~\cite{Sood:PhysiaA_224:1996,Strepp:PRE_63:2001,Chaudhuri:PRE_72:2005}.

The set-up of a modulated liquid adds two more control parameters with respect to the same bulk liquid. The corresponding intermediate scattering functions and derived quantities provide relevant information on the competition of local packing and the imposed modulation. Depending on the period and strength of the modulation a non-trivial nonequilibrium state diagram follows already for the simplest conceivable interaction, the hard-sphere potential. In the companion paper~\cite{Ahmadirahmat:PRE:2025} we  evaluate the theory numerically for the nonergodicity parameters for the case of a monolayer of hard-sphere particles  and determine the glass-transition lines and show a reentrant behavior emerges.  We anticipate that upon tuning the parameters also higher-order singularities emerge similar to the colloid-polymer mixtures~\cite{Bergenholtz:PRE_59:1999,Fabbian:PRE_59:1999,Dawson:PRE_63:2000,Pham:Science_296:2002} or liquids in a porous host matrix~\cite{Krakoviack:PRL_94:2005, *Krakoviack:PRE_75:2007, *Krakoviack:PRE_79:2009,Kurzidem:PRL_103:2009,Kim:EPL_88:2009}.

The most promising realization of a modulated liquid consists of a colloidal monolayer where the external periodic potential  is achieved in terms of interfering laser beams~\cite{Dalle-Ferrier:SM_7:2011, Jenkins:JoP_40:2008,Capellmann:JCP_148:2018}. Our theory should directly apply to this situation, a comparison to experiments is the content of future work.

\acknowledgements
We thank Rolf Schilling for constructive criticism on the manuscript. This research was funded in part by the Austrian Science Fund (FWF) 10.55776/I5257 and 10.55776/P35872.

\appendix

\section{Lattice sums}\label{Sec:lattice_sums} 
In this appendix we collect some properties of discrete Fourier transforms in periodic systems, thereby introducing some useful notation. 

The volume is considered to be a hypercube $V=L^d$ of linear dimension $L$. In particular, the box size is an integer multiple of the period of the potential, $L = M a$ with $M\in \mathbb{N}$. All functions are considered to be periodically extended 
$f(\vec{x}) = f(\vec{x}+ \vec{R}_\lambda)$ where $\vec{R}_\lambda \in \lambda:= \{ \vec{R} \in \mathbb{R}^d : \vec{R} \in ( L \mathbb{Z})^d \}$. Such functions will be referred to as $\lambda$-periodic. 
Then Fourier modes are introduced according to 
\begin{align}\label{eq:lattice_Fourier}
\hat{f}(\vec{k})  =  \int_V \diff \vec{x} f(\vec{x}) \exp(i \vec{k} \cdot \vec{x} )  .
\end{align} 
Since $f(\vec{x})$ is $\lambda$-periodic, a shift $\vec{x}\mapsto \vec{x} + \vec{R}_\lambda$ with $\vec{R}_\lambda\in \lambda$ leaves the function invariant, and we have to impose
\begin{align}
\exp(i \vec{k} \cdot \vec{R}_\lambda ) = 1 \qquad \text{for }   \vec{R}_\lambda\in \lambda,
\end{align}
which restricts the wave vectors to the \emph{reciprocal lattice} $\vec{k}\in \lambda^* := \{ \vec{k} \in \mathbb{R}^d : \vec{k} \in (2\pi \mathbb{Z} /L)^d \}$. 
The following relations are useful
\begin{subequations}
\begin{align}
\frac{1}{V} \int_V \diff \vec{x} \exp( i \vec{k} \cdot \vec{x} ) &=  \delta_{\vec{k},0} \qquad 
\text{for } \vec{k}\in \lambda^* ,\\
\frac{1}{V} \sum_{\vec{k} \in \lambda^*} \exp( -i \vec{k} \cdot \vec{x} ) &= \delta(\vec{x}) \qquad \text{for } \vec{x}\in V.  
\end{align}
\end{subequations}
Then we find also the representation of a $\lambda$-periodic function in terms of its Fourier modes
\begin{align}
f(\vec{x}) = \frac{1}{V} \sum_{\vec{k} \in \lambda^*} \hat{f}(\vec{k}) \exp(-i \vec{k}\cdot \vec{x}) .
\end{align}

Now for modulated liquids we decompose $\vec{k} = \vec{q} + \vec{Q}_\mu$ such that $\vec{q}\in \text{BZ} := \{ \vec{q} \in \lambda^*: -\pi/a < \vec{q}\cdot\vec{e}_z \leq \pi /a\} , \vec{Q}_\mu \in \Lambda^* = \{ \vec{Q}_\mu = (2\pi\mu/a) \vec{e}_z : \mu\in \mathbb{Z} \}$ and use the compact notation $\vec{q}_\mu = \vec{q}+\vec{Q}_\mu$ to indicate the decomposition. The corresponding Fourier mode is abbreviated as $\hat{f}_\mu(\vec{q}) := \hat{f}(\vec{q}_\mu)$. Then the Fourier decomposition can be expressed as
\begin{subequations}
\begin{align}
\hat{f}_\mu(\vec{q}) &=  \int_V \diff \vec{x}\, f(\vec{x}) \exp[i (\vec{q}+\vec{Q}_\mu) \cdot \vec{x} ) ], \\
f(\vec{x}) &= \frac{1}{V} \sum_{\mu\in \mathbb{Z}} \sum_{\vec{q} \in\text{BZ}} \hat{f}_\mu(\vec{q}) \exp[-i (\vec{q} + \vec{Q}_\mu ) \cdot \vec{x} ]  .
\end{align}
\end{subequations}
The orthogonality and completeness is encoded in
\begin{subequations}
\begin{align}\label{eq:orthogonality_completeness}
\frac{1}{V} \int \diff \vec{x} \exp[ i(\vec{q}+\vec{Q}_\mu) \cdot \vec{x} ] &= \delta_{\vec{q},0}  \delta_{\mu,0} 
, \\
\frac{1}{V} \sum_{\mu\in\mathbb{Z}} \sum_{\vec{q} \in \text{BZ}} \exp[-i (\vec{q}+\vec{Q}_\mu ) \cdot \vec{x} ] &= \delta(\vec{x})  .
\end{align}
\end{subequations}
 where $\vec{q} \in \text{BZ}, \vec{Q}_\mu \in \Lambda^*, \vec{x} \in V$.  
 
 It is also of interest to specialize the decomposition for functions that are already $\Lambda$-periodic, $\mathcal{U}(\vec{x}) = \mathcal{U}(\vec{x}+\vec{R})$ with $\vec{R}\in \Lambda = \{ \vec{r} + n a \vec{e}_z: \vec{r} \perp \vec{e}_z, n \in \mathbb{Z} \}$. Then 
 \begin{align}\label{eq:Fourier_periodic1}
 \hat{\mathcal{U}}_\mu(\vec{q}) = \delta_{\vec{q},0} V \frac{1}{a} \int_0^a \diff z \, \mathcal{U}(z) e^{i Q_\mu z} = \delta_{\vec{q},0} V \,\mathcal{U}_\mu,
 \end{align}
 where $\mathcal{U}_\mu$ is the Fourier coefficient of the one-dimensional modulation as defined in Eq.~\eqref{eq:def_Fourier}.

\section{Static current-density correlator}\label{Sec:Static_current}

In this appendix we calculate the thermal average of the current density correlation function that is needed for introducing the projector on the currents. 

The static current density correlation matrix $\mathcal{K}^{\alpha\beta}_{\mu\nu}(\vec{q}) = N^{-1}\langle j^\alpha_\mu(\vec{q})^* j^\beta_\nu(\vec{q}) \rangle $  is diagonal with respect to the channel indices $\alpha,\beta$ since averages over unpaired momenta vanish. Inserting the current densities and pre-averaging over the momenta yields
\begin{align}\label{eq:static_current_matrix}
\mathcal{K}^{\alpha\beta}_{\mu\nu}(\vec{q}) =&\sum_{n,m=1}^N \Big\langle \langle b^{\alpha}[  (\hat{\vec{q}}^\parallel\cdot \vec{p}_n ) (\hat{\vec{q}}^\parallel\cdot \vec{p}_{m} ), (\hat{\vec{q}}^\perp \cdot \vec{p}_n) (\hat{\vec{q}}^\perp \cdot \vec{p}_{m} )  ] \rangle \nonumber \\
& \times  \exp[-i (\vec{q}+\vec{Q}_\mu ) \cdot \vec{x}_{m} ]
 \exp[i (\vec{q}+\vec{Q}_\nu ) \cdot \vec{x}_n ] \Big\rangle   \frac{ \delta_{\alpha\beta} }{N m^2}.
\end{align}

The thermal averages of the momenta follow from the equipartition theorem 
\begin{align}
\langle (\hat{\vec{q}}^\parallel\cdot \vec{p}_n ) (\hat{\vec{q}}^\parallel\cdot \vec{p}_{m} )  \rangle = \langle (\hat{\vec{q}}^\perp\cdot \vec{p}_n )(\hat{\vec{q}}^\perp\cdot \vec{p}_{m} ) \rangle = \delta_{n m} m k_B T .
\end{align}
Then 
\begin{align}
\mathcal{K}^{\alpha\beta}_{\mu\nu}(\vec{q}) &= \frac{k_B T}{N m } \delta^{\alpha\beta} \sum_{n=1}^N \langle \exp[ -i (\vec{Q}_\mu-\vec{Q}_\nu) \cdot \vec{x}_n ] \rangle 
\nonumber \\
&= \frac{k_B T}{N m}\delta^{\alpha\beta} \int_V \diff \vec{x} \sum_{n=1}^N \langle \delta(\vec{x}-\vec{x}_n) \rangle   \exp[ -i \vec{Q}_{\mu-\nu} \cdot \vec{x} ] \nonumber \\
&= \frac{k_B T}{N m}\delta^{\alpha\beta} \int_V \diff \vec{x}\, \, n(z) \exp[ -i Q_{\mu-\nu} {z} ] \nonumber \\
&= \frac{k_B T}{N m}\delta^{\alpha\beta}   \frac{V}{L} \int_0^L \diff z \, \, n(z) \exp[ -i Q_{\mu-\nu}{z} ] \nonumber \\
\mathcal{K}^{\alpha\beta}_{\mu\nu}(\vec{q}) &= \frac{k_B T}{m n_{0}} \delta^{\alpha\beta} n_{\nu-\mu} .
\end{align}
Note that the correlation functions for the split currents is expressible solely in terms of the density profile.

\section{Ornstein-Zernike equation}\label{sec:Ornstein_Zernike}

In this appendix we recall the definition of some static correlation functions in inhomogeneous liquids and specialize to the case of a modulated liquid. 
 
The total correlation function $h(\vec{x}_1,\vec{x}_2)$ is defined~\cite{Hansen:Theory_of_Simple_Liquids} via the 
 initial value of $G(\vec{x}_1,\vec{x}_2)= G(\vec{x}_1,\vec{x}_2,t=0)$ of the van Hove correlation function by
\begin{align}
 n_{0} G(\vec{x}_1,\vec{x}_2) = n(\vec{x}_1) h(\vec{x}_1,\vec{x}_2) n(\vec{x}_2) + n(\vec{x}_1) \delta(\vec{x}_1-\vec{x}_2).
\end{align}
 Similarly the direct correlation function is defined implicitly via the Ornstein-Zernike relation~\cite{Hansen:Theory_of_Simple_Liquids} 
\begin{align}
 h(\vec{x}_1,\vec{x}_2) =& c(\vec{x}_1,\vec{x}_2) \nonumber \\
& + \int_V c(\vec{x}_1,\vec{x}_3) n(\vec{x}_3) h(\vec{x}_3,\vec{x}_2) \diff \vec{x}_3 ,
\end{align}
which implies
\begin{align}
 n_{0} G(\vec{x}_1,\vec{x}_2) =& n(\vec{x}_1) \delta(\vec{x}_1-\vec{x}_2) \nonumber \\
&+ n(\vec{x}_1) \int_V c(\vec{x}_1,\vec{x}_3) n_{0} G(\vec{x}_3,\vec{x}_2)\diff \vec{x}_3 ,
\end{align}
or equivalently 
\begin{equation}
  v(\vec{x}_1) n_{0} G(\vec{x}_1,\vec{x}_2) =  \delta(\vec{x}_1-\vec{x}_2) +  \int c(\vec{x}_1,\vec{x}_3) n_{0} G(\vec{x}_3,\vec{x}_2)\diff \vec{x}_3 . 
\end{equation}

Let's decompose the  terms into Fourier modes relying on the representation formulas, Eqs. \eqref{eq:representation_1},\eqref{eq:representation_2}  and the orthogonality relations, Eq. \eqref{eq:orthogonality_completeness}. 
For the first term we calculate using Eqs.\eqref{eq:periodic},~\eqref{eq:representation_1}
\begin{align}
&\frac{1}{V}\int_V\!\! \diff \vec{x}_1 \!\int_V\!\! \diff \vec{x}_2 e^{-i \vec{q}_\mu \cdot \vec{x}_1}  
v(\vec{x}_1) n_{0} G(\vec{x}_1,\vec{x}_2) 
e^{i\vec{q}_\nu \cdot \vec{x}_2} = \nonumber \\
& \frac{1}{V}\int_V\!\! \diff \vec{x}_1\! \int_V\!\! \diff \vec{x}_2 e^{-i \vec{q}_\mu \cdot \vec{x}_1}  
\sum_{\kappa\in \mathbb{Z} } v_{-\kappa} e^{i \vec{Q}_\kappa \cdot\vec{x}_1 } 
 n_{0} G(\vec{x}_1,\vec{x}_2) 
e^{i\vec{q}_\nu  \cdot \vec{x}_2} \nonumber \\
&= n_{0}\sum_{\kappa\in \mathbb{Z} } v_{-\kappa}  S_{\mu-\kappa, \nu}(\vec{q}) .
\end{align}
The second terms yields 
\begin{align}
&\frac{1}{V}\int_V\!\! \diff \vec{x}_1\! \int_V\! \diff \vec{x}_2 e^{-i \vec{q}_\mu \cdot \vec{x}_1}  
\delta(\vec{x}_1-\vec{x}_2) 
e^{i \vec{q}_\nu  \cdot \vec{x}_2} = \nonumber \\
&= \frac{1}{V} \int_V\!\! \diff \vec{x}_1 e^{-i \vec{q}_\mu \cdot \vec{x}_1}   
e^{i \vec{q}_\nu  \cdot \vec{x}_1} = \delta_{\mu\nu} .
\end{align}
And for the last term
\begin{align}
&\frac{1}{V}\int_V\!\! \diff \vec{x}_1\! \int_V\!\! \diff \vec{x}_2 e^{-i \vec{q}_\mu \cdot \vec{x}_1}  
\int_V\!\! \diff \vec{x}_3  c(\vec{x}_1,\vec{x}_3) n_{0} G(\vec{x}_3,\vec{x}_2 )
  e^{i\vec{q}_\nu  \cdot \vec{x}_2}  \nonumber \\
 =& \frac{1}{V} \int_V\!\! \diff \vec{x}_1 \! \int_V\!\! \diff \vec{x}_2\! \int_V\!\! \diff \vec{x}_3 e^{-i \vec{q}_\mu \cdot \vec{x}_1}
 \frac{1}{V} \sum_{\lambda,\kappa\in \mathbb{Z} } \sum_{\vec{k}\in \text{BZ}} c_{\lambda\kappa}(\vec{k}) \nonumber \\
&\times e^{i\vec{k}_\lambda\cdot \vec{x}_1 } e^{-i \vec{k}_\kappa \cdot \vec{x}_3 } n_{0} G(\vec{x}_3 , \vec{x}_2 ) e^{i \vec{q}_\nu \cdot \vec{x}_2 } \nonumber \\
 =& \int_V\!\! \diff \vec{x}_2 \!\int_V\!\! \diff \vec{x}_3 
 \frac{1}{V} \sum_{\lambda,\kappa\in \mathbb{Z} } \sum_{\vec{k}\in \text{BZ}} \delta _{\mu\lambda} 
\delta_{\vec{q},\vec{k} } c_{\lambda\kappa}(\vec{k})  e^{-i \vec{k}_\kappa \cdot \vec{x}_3 } \nonumber\\ 
&\times n_{0} G(\vec{x}_3 , \vec{x}_2 ) e^{i \vec{q}_\nu \cdot \vec{x}_2 } \nonumber \\
 =& \int_V\!\! \diff \vec{x}_2 \!\int_V\!\! \diff \vec{x}_3 
 \frac{1}{V} \sum_{\kappa\in \mathbb{Z} } c_{\mu\kappa}(\vec{q})  e^{-i \vec{q}_\kappa \cdot \vec{x}_3 } n_{0} G(\vec{x}_3 , \vec{x}_2 ) e^{i \vec{q}_\nu \cdot \vec{x}_2 } \nonumber \\
 =& n_{0}\sum_{\kappa\in \mathbb{Z}} c_{\mu\kappa}(\vec{q})  S_{\kappa\nu}(\vec{q}) .
\end{align} 
After collecting results and relabeling indices we find 
\begin{align}
n_{0} \sum_{\kappa \in \mathbb{Z}}   [v_{\kappa-\mu}- c_{\mu\kappa}(\vec{q}) ]  S_{\kappa\nu}(\vec{q})   = \delta_{\mu\nu} ,
\end{align}
or equivalently
\begin{align}
[\mathbf{S}^{-1}(\vec{q}) ]_{\mu\nu} = n_{0} v_{ \nu-\mu} - n_{0} c_{\mu\nu}(\vec{q})  ,
\end{align}
which is Eq.~\eqref{eq:Ornstein-Zernike} of the main text.

\section{Evaluation of  the overlap matrix element}\label{sec:scalarproduct_a} 
In this appendix  we calculate the scalar product
$\langle [ \mathcal{Q} \mathcal{L}j_{\mu}^{\alpha}(\vec{q})]^*\delta\rho_{\mu_1}(\vec{q_{1}})\delta\rho_{\mu_2}(\vec{q}_2)\rangle$ required for the mode-coupling vertex in Eq.~\eqref{eq:overlap}. 

With $\mathcal{Q}=1-\mathcal{P}_{j}-\mathcal{P}_{\rho}$ and $\mathcal{P}_{j} |\delta\rho_{\mu_1}(\vec{q_{1}})\delta\rho_{\mu_2}(\vec{q}_2)\rangle=0$ by time inversion symmetry, one obtains three contributions
\begin{align}\label{eq:three}
&\langle [\mathcal{Q} \mathcal{L}j_{\mu}^{\alpha}(\vec{q})]^*\delta\rho_{\mu_1}(\vec{q_{1}})\delta\rho_{\mu_2}(\vec{q}_2)\rangle\nonumber \\
=&\langle [j_{\mu}^{\alpha}(\vec{q})]^*[\mathcal{L}\delta\rho_{\mu_1}(\vec{q_{1}})]\delta\rho_{\mu_2}(\vec{q}_2)\rangle+(1\leftrightarrow2)\nonumber \\
&-\langle [\mathcal{L}j_{\mu}^{\alpha}(\vec{q})]^*\mathcal{P}_{\rho} [\delta \rho_{\mu_1}(\vec{q_{1}})\delta\rho_{\mu_2}(\vec{q}_2)]\rangle.
\end{align}
For the  first term the particle-conservation law  Eq.~(\ref{eq:continuity}) implies
\begin{align}\label{eq:scalar}
& \langle [j_{\mu}^{\alpha}(\vec{q})]^*[\mathcal{L}\delta\rho_{\mu_1}(\vec{q_{1}})]\delta\rho_{\mu_2}(\vec{q}_2)\rangle=\nonumber\\
&\sum_{\gamma=\parallel,\perp} (q_1+Q_{\mu_1})^\gamma
\langle j_{\mu}^{\alpha}(\vec{q})^*j_{\mu_{1}}^{\gamma}(\vec{q}_{1})\delta \rho_{\mu_{2}}(\vec{q}_{2})\rangle.
\end{align}
Again, averaging over the momenta first, and then over the positions similar to Eq.~(\ref{eq:static_current_matrix}),
one obtains
\begin{align}
&\langle j_{\mu}^{\alpha}(\vec{q})^*j_{\mu_{1}}^{\gamma}(\vec{q}_{1}) \delta\rho_{\mu_{2}}(\vec{q}_{2})\rangle\nonumber\\
&=\delta^{\alpha\gamma} v_\text{th}^2 
\langle\rho(\vec{q}-\vec{q}_1 + \vec{Q}_\mu - \vec{Q}_{\mu_1} )|\rho(\vec{q}_{2}+ \vec{Q}_{\mu_2})\rangle. 
\end{align}
where we temporarily reintroduced the fluctuating density $\rho(\vec{k})$ to total wavevector $\vec{k}$. 
By discrete translational symmetry this matrix element is only non-vanishing if the crystal momentum is conserved $\vec{q}-\vec{q}_1- \vec{q}_2 \in \Lambda^*$. In our case of a one-dimensional modulation we need to consider only the cases $\vec{q}-\vec{q}_1-\vec{q}_2= 0$ and $\vec{q}-\vec{q}_1-\vec{q}_2= \vec{Q}_{\pm 1}$, the first case is referred to as a \emph{normal scattering} (N-process) while the latter are called \emph{umklapp}-processes (U-processes).  
Hence we can formally combine the expression to 
\begin{align}
\lefteqn{ \langle j_{\mu}^{\alpha}(\vec{q})^*j_{\mu_{1}}^{\gamma}(\vec{q}_{1}) \delta\rho_{\mu_{2}}(\vec{q}_{2})\rangle } \nonumber \\ &
=\delta^{\alpha\gamma} N v_\text{th}^2  \sum_{\mu^\sharp =0,\pm 1} \delta_{\vec{q}-\vec{Q}_{\mu^\sharp},\vec{q}_1+ \vec{q}_2}  
S_{\mu-\mu_{1}+\mu^\sharp,\mu_{2}}(\vec{q}_{2}),
\end{align}
where at most one of the terms $\mu^\sharp = 0,\pm 1$  contributes. 

Evaluating the projection on the density modes in  the third term in Eq.~(\ref{eq:three})   leads to
\begin{align}
&\langle[\mathcal{L}j_{\mu}^{\alpha}(\vec{q})]^* {\cal P}_{\rho}[ \delta\rho_{\mu_1}(\vec{q_{1}})\delta\rho_{\mu_2}(\vec{q}_2) ]\rangle \nonumber\\
=&\frac{1}{N}\sum_{\mu',\mu'' \in \mathbb{Z}}\langle j_{\mu}^{\alpha}(\vec{q})|\mathcal{L}\rho_{\mu'}(\vec{q})\rangle [\mathbf{S}^{-1}(\vec{q})]_{\mu'\mu''} \nonumber\\
&\times\langle\delta\rho_{\mu''}(\vec{q})^*\delta\rho_{\mu_{1}}
(\vec{q}_{1})\delta\rho_{\mu_{2}}(\vec{q}_{2})\rangle\nonumber\\
=&\frac{1}{N}\sum_{\mu',\mu'' \in\mathbb{Z}}
\sum_{\beta=\parallel, \perp} (q+Q_{\mu'})^\beta
\langle j_{\mu}^{\alpha}(\vec{q})|j_{\mu'}^{\beta}(\vec{q})\rangle [\mathbf{S}^{-1}(\vec{q})]_{\mu'\mu''} \nonumber\\
&\times\langle\delta\rho_{\mu''}(\vec{q})^*\delta\rho_{\mu_{1}}
(\vec{q}_{1})\delta\rho_{\mu_{2}}(\vec{q}_{2})\rangle,
\end{align}
where  the particle-conservation law,  Eq.~\eqref{eq:continuity}, has been used again. Substituting Eq.~(\ref{eq:static_current_matrix}) for the current-current static correlator
  the projected matrix element evaluates to
\begin{align}
\lefteqn{ \langle\mathcal{L}j_{\mu}^{\alpha}(\vec{q})|\mathcal{P}_{\rho}|\delta\rho_{\mu_1}(\vec{q_{1}})\delta\rho_{\mu_2}(\vec{q}_2)\rangle =} \nonumber\\
=&
N v_\text{th}^2 
\sum_{\mu',\mu''\in \mathbb{Z}}
\frac{n_{\mu'-\mu}}{n_{0}}
(q+Q_{\mu'})^\alpha \nonumber\\
&\times[\mathbf{S}^{-1}(\vec{q})]_{\mu'\mu''}S_{\mu'',\mu_{1}\mu_{2}}(\vec{q},\vec{q}_{1}\vec{q}_{2}).
\end{align}
Here, we abbreviated
the static three-point correlation function by $S_{\mu,\mu_{1}\mu_{2}}(\vec{q},\vec{q}_{1}\vec{q}_{2})= N^{-1} \langle\delta\rho_{\mu}(\vec{q})^*\delta\rho_{\mu_{1}}
(\vec{q}_{1})\delta\rho_{\mu_{2}}(\vec{q}_{2})\rangle$. Note that the triple correlation function is non-vanishing only if $\vec{q}-\vec{q}_1-\vec{q}_2 \in \Lambda^*$. 
Collecting terms one finds Eq.~(\ref{eq:overlap}) of the main text
\begin{align}
&\langle\mathcal{Q} \mathcal{L}j_{\mu}^{\alpha}(\vec{q})|\delta\rho_{\mu_1}(\vec{q}_{1})\delta\rho_{\mu_2}(\vec{q}_2)\rangle\nonumber\\
=&N v_\text{th}^2 \sum_{\mu^\sharp=0,\pm 1} \delta_{\vec{q}-\vec{Q}_{\mu^\sharp},\vec{q}_{1}+\vec{q}_{2}}\big\{
(q_1+ Q_{\mu_1})^\alpha
S_{\mu-\mu_{1}+\mu^\sharp,\mu_{2}}(\vec{q}_{2})\nonumber \\
&+(1\leftrightarrow2) -\sum_{\mu',\mu'' \in \mathbb{Z}} \frac{n_{\mu'-\mu}}{n_{0}}
(q+Q_{\mu'})^\alpha
 [\mathbf{S}^{-1}(\vec{q})]_{\mu'\mu''} \nonumber \\
&\times S_{\mu'',\mu_{1}\mu_{2}}(\vec{q},\vec{q}_{1}\vec{q}_{2})\big\}.
\end{align}

\section{Vertex approximation}
\label{sec:vertex_approximation}

In this appendix, we complete the calculation of the MCT vertex, using
the convolution approximation in order to express the static
three-point correlation function in terms of products of two-point
correlation functions.

The vertex after evaluating the overlap matrix elements is given by
three terms:
\begin{align}\label{eq:vertex_long_calculation}
\lefteqn{ \mathcal{X}^{\alpha}_{\mu,\mu_{1}\mu_{2}}(\vec{q},\vec{q}_{1}\vec{q}_{2}) = 
 N v_\text{th}^2 \sum_{\mu^\sharp=0,\pm 1} \delta_{\vec{q}-\vec{Q}_{\mu^\sharp},\vec{q}_{1}+\vec{q}_{2}} 
\Big\{  } \nonumber \\
& \sum_{\mu_1',\mu_2' \in \mathbb{Z} }
\Big [(q_1+ Q_{\mu_1'})^\alpha
S_{\mu-\mu_{1}'+\mu^\sharp,\mu_{2}'}(\vec{q}_{2})  [\mathbf{S}^{-1}(\vec{q}_1) ]_{\mu_1',\mu_1} [\mathbf{S}^{-1}(\vec{q}_2)]_{\mu_2'\mu_2}\nonumber \\
&+(1\leftrightarrow2)  \Big] -  \sum_{\mu_1'\mu_2' \in \mathbb{Z} } \sum_{\mu'\mu'' \in \mathbb{Z}} \frac{n_{\mu'-\mu}}{n_{0}}
(q+Q_{\mu'})^\alpha
 [\mathbf{S}^{-1}(\vec{q})]_{\mu'\mu''} \nonumber \\
&\times S_{\mu'',\mu_{1}'\mu_{2}'}(\vec{q},\vec{q}_{1}\vec{q}_{2}) [\mathbf{S}^{-1}(\vec{q}_1) ]_{\mu_1'\mu_1} [\mathbf{S}^{-1}(\vec{q}_2)]_{\mu_2'\mu_2} \Big\}  .
\end{align}
For the first two terms in the curly bracket, the sums over
$(\mu_{1}',\mu_{2}')$ can be performed which leads to
\begin{align}
& (q_1+Q_{\mu+\mu^\sharp-\mu_2})^\alpha
  [\mathbf{S}^{-1}(\vec{q}_{1})]_{\mu+\mu^\sharp-\mu_{2},\mu_{1}} \nonumber \\
 & + 
(q_2+Q_{\mu+\mu^\sharp-\mu_1})^\alpha
  [\mathbf{S}^{-1}(\vec{q}_{2})]_{\mu+\mu^\sharp-\mu_{1},\mu_{2}}.
\end{align}
Inserting the Ornstein-Zernike equation, Eq.~\eqref{eq:Ornstein-Zernike}, they can
be recast to
\begin{align}\label{eq:firsttwoterms}
& n_{0} \Big[ (q_1+q_2 + 2 Q_{\mu+\mu^\sharp} - Q_{\mu_1}-Q_{\mu_2} )^\alpha  v_{\mu_1+\mu_2-\mu-\mu^\sharp} \nonumber \\
&- (q_1+ Q_{\mu+\mu^\sharp-\mu_2} )^\alpha  c_{\mu+\mu^\sharp-\mu_2,\mu_1}(\vec{q}_1) \nonumber \\
&- (q_2+ Q_{\mu+\mu^\sharp-\mu_1} )^\alpha  c_{\mu+\mu^\sharp-\mu_1,\mu_2}(\vec{q}_2)  \Big].
\end{align}
 As for the third term, the convolution approximation (see
Appendix \ref{sec:convolution2}) yields for the triplet structure
factor 
\begin{align} 
 S_{\mu'',\mu_1' \mu_2'} & (\vec{q},\vec{q}_1\vec{q}_2)    
 \approx n_{0}^2 \sum_{\nu^\sharp = 0,\pm 1}  \delta_{\vec{q}-\vec{Q}_{\nu^\sharp}, \vec{q}_1+ \vec{q}_2 }  \nonumber \\
 & \times \sum_{ \substack{\nu', \nu_1', \nu_2' \in \mathbb{Z} \\ \nu'', \nu_1'', \nu_2'' \in \mathbb{Z} }} 
n_{\nu' + \nu_1'+ \nu_2' - \nu^\sharp}  v_{-\nu'-\nu''} S_{\mu'',\nu'',}(\vec{q})\nonumber \\
& \times v_{-\nu_1'+ \nu_1''} S_{\nu_1'', \mu_1'}(\vec{q}_1)  
v_{-\nu_2'+\nu_2''} S_{\nu_2'', \mu_2'}(\vec{q}_2)  . 
\end{align}
The sum over $\nu^\sharp$  enforces $\nu^\sharp = \mu^\sharp$ once we consider the outer sum in 
Eq.~\eqref{eq:vertex_long_calculation}.  One can then  perform the sums over $\mu'', \mu_1', \mu_2'$  contracting all structure factors with their corresponding inverses. The resulting Kronecker symbols allow performing the sums over $\nu'', \nu_1'', \nu_2''$. 
The convolution theorem for the densities, Eq.~\eqref{eq:convolution_theorem}, permits performing the sum over $\nu'$ and then successively over $\mu'$ yielding as result for the third term in the curly bracket
\begin{align}
- n_{0} \sum_{\nu_1', \nu_2'\in \mathbb{Z} } n_{\nu_1'+\nu_2'- \mu- \mu^\sharp} (q+ Q_{\nu_1'+\nu_2'-\mu^\sharp})^\alpha v_{\mu_1-\nu_1'} v_{\mu_2-\nu_2'}  .
\end{align}
Eventually, splitting the sum in three terms according to
$(q+Q_{\nu_1'+\nu_2'-\mu^\sharp})^\alpha = (q- Q_{\mu^\sharp})^\alpha + Q_{\nu_1'}^\alpha + Q_{\nu_2'}^\alpha$, one can
perform the last summations over $\nu_1'$ and $\nu_2'$ (in a
suitable order depending on the presence of  $Q_{\nu_1'}^\alpha$ or
$Q_{\nu_2'}^\alpha$), and the third term in Eq.~\eqref{eq:vertex_long_calculation} in the curly bracket  reduces
to
\begin{align}
- n_{0}( q+ Q_{2 \mu + \mu^\sharp- \mu_1-\mu_2})^\alpha v_{\mu_1+\mu_2-\mu-\mu^\sharp}.
\end{align}
This contribution  precisely cancels the first line in Eq.~\eqref{eq:firsttwoterms} since $\vec{q} = \vec{q}_1+ \vec{q}_2 +\vec{ Q}_{\mu^\sharp}$. Collecting terms, the vertex thus simplifies to
\begin{align}
&  \mathcal{X}^{\alpha}_{\mu,\mu_{1}\mu_{2}}
  (\vec{q},\vec{q}_{1}\vec{q}_{2})  \approx -N n_{0} v_\text{th}^2 \sum_{\mu^\sharp = 0, \pm 1} 
  \delta_{\vec{q}-\vec{Q}_{\mu^\sharp} ,\vec{q}_{1}+\vec{q}_{2}}  \nonumber \\
  \times &  \left[ (q_1+ Q_{\mu+\mu^\sharp}-Q_{\mu_2})^\alpha
    c_{\mu+ \mu^\sharp-\mu_2,\mu_1}(\vec{q}_1) \right.  \nonumber \\
  & \left. + (q_2+ Q_{\mu+\mu^\sharp} -Q_{\mu_1})^\alpha
  c_{\mu+\mu^\sharp-\mu_1,\mu_2}(\vec{q}_2) \right], 
\end{align}
which is of a similar  form as for simple \cite{Goetze:Complex_Dynamics}
and molecular liquids \cite{Scheidsteger:PRE56:1997,Fabbian:PRE_59:1999}.

\section{Convolution approximation}
\label{sec:convolution2}

In this appendix, the convolution approximation for the triplet
structure factor of a fluid exposed to  a periodic potential energy
landscape is derived.  The procedure is based on the general
formulation of Rajan \emph{et al.} \cite{Rajan:JMath19:1978}, valid for any
inhomogeneous system, and is similar to the one used in
Ref.~\cite{Lang:PRE_86:2012} for a fluid in a slit pore.

We first consider the general case of an inhomogeneous $N$-particle
fluid enclosed in a $d$-dimensional hypercube of volume
$V$, for which we lay down the necessary definitions and
notations.  With
\begin{equation}
  \rho(\vec{k}) = \sum_{l=1}^N e^{i\vec{k}\cdot\vec{x}_l},
\end{equation}
where $\vec{x}_n$ denotes the position of the $n$-th particle and $\vec{k}\in \lambda^*$ is a wave vector compatible with the finite box, see Appendix~\ref{Sec:lattice_sums},
the
Fourier transform of the one-body density, denoted with $n(\vec{x})$,
is given by
\begin{equation}
 \hat{n}(\vec{k}) = \langle \rho(\vec{k}) \rangle.
\end{equation}
As for the pair and triplet structure factors, they are defined from
the fluctuations
\begin{equation}
  \delta\rho(\vec{k}) = \rho(\vec{k}) - \langle \rho(\vec{k}) \rangle
  = \rho(\vec{k}) - \hat{n}(\vec{k}), 
\end{equation}
as
\begin{align}
  S^{(2)}({\vec{k}_0,\vec{k}_{1}}) = \frac{1}{N} \langle
  \delta\rho(\vec{k}_0) \delta\rho(\vec{k}_{1}) \rangle, \\
  S^{(3)}({\vec{k}_0,\vec{k}_{1},\vec{k}_{2}}) = \frac{1}{N} \langle
  \delta\rho(\vec{k}_0) \delta\rho(\vec{k}_{1})
  \delta\rho(\vec{k}_{2}) \rangle,
\end{align}
respectively.  Note that, at variance with the main text, the
definitions of the structure factors in this Appendix do not involve
any complex conjugation in order to preserve the symmetry of the
working equations.  It is also convenient to introduce the Fourier
components $\hat{v}(\vec{k})$, Eq.~\eqref{eq:lattice_Fourier}, of the local specific volume
$v(\vec{x}) = 1/n(\vec{x})$, which obey generally
\begin{equation}
  \frac{1}{V} \sum_{\vec{k}} \hat{n}(\vec{k}_1-\vec{k})
  \hat{v}(\vec{k}-\vec{k}_2) = V \delta_{\vec{k}_1,\vec{k}_2}.
\end{equation}
Then, as shown in Ref.~\cite{Lang:PRE_86:2012}, the convolution approximation
for $S^{(3)}({\vec{k}_0,\vec{k}_{1},\vec{k}_{2}})$ obtained in
Ref.~\cite{Rajan:JMath19:1978} can be written as
\begin{align} \label{eq:convolution} %
&  S^{(3)}({\vec{k}_0,\vec{k}_{1},\vec{k}_{2}}) \approx
  \frac{N^2}{V^6} \nonumber \\
  &\times \sum_{\substack{\vec{k}'_0,\vec{k}'_1,\vec{k}'_2
      \\ \vec{k}''_0,\vec{k}''_1,\vec{k}''_2}} \hat{n}(\vec{k}'_0 +
  \vec{k}'_1 + \vec{k}'_2) \prod_{i=0}^2
  \hat{v}(-\vec{k}'_i-\vec{k}''_i) S^{(2)}(\vec{k}''_i,\vec{k}_i).
\end{align}
Specialization to the case of a fluid in a periodic modulation is
almost straightforward.  Here we present the derivation for the slightly more general case of an arbitrary Bravais lattice $\Lambda$ with associated reciprocal lattice $\Lambda^*$ and (first) Brillouin zone $\text{BZ}$.   
First, a generic  wave vector $\vec{k} \in \lambda^*$, see Appendix~\ref{Sec:lattice_sums}, uniquely decomposes into $\vec{k}= \vec{q}+\vec{Q}$ where $\vec{Q}\in \Lambda^*$  belongs
to the reciprocal lattice, and $\vec{q}\in \text{BZ}$, to the first Brillouin zone.  Accordingly, the sums
over wave vectors in Eq.~\eqref{eq:convolution} split into separate
sums over each type of wave vectors,
\begin{equation}
   \sum_{\substack{\vec{k}'_0,\vec{k}'_1,\vec{k}'_2
       \\ \vec{k}''_0,\vec{k}''_1,\vec{k}''_2}} \ldots =
   \sum_{\substack{\vec{Q}'_0,\vec{Q}'_1,\vec{Q}'_2 \in \Lambda^*
       \\ \vec{Q}''_0,\vec{Q}''_1,\vec{Q}''_2 \in \Lambda^*}}
   \sum_{\substack{\vec{q}'_0,\vec{q}'_1,\vec{q}'_2 \in \text{BZ}
       \\ \vec{q}''_0,\vec{q}''_1,\vec{q}''_2 \in \text{BZ}}} \ldots.
\end{equation}
Second, using the decomposition we write $S^{(2)}_{\vec{Q}_0 \vec{Q}_1}(\vec{q}_0, \vec{q}_1) :=  S^{(2)}(\vec{k}_0,\vec{k}_1)$, and 
as a result of lattice periodicity this is non-vanishing only if  $\vec{k}_0+\vec{k}_1 \in \Lambda^*$ is a reciprocal lattice vector or equivalently if $\vec{q}_0 +\vec{q}_1=0$.  Therefore we conclude
\begin{equation}
S^{(2)}(\vec{k}_0,\vec{k}_1) =
  S^{(2)}_{\vec{Q}_0\vec{Q}_1}(-\vec{q}_1,\vec{q}_1)
  \delta_{\vec{q}_0+\vec{q}_1,\vec{0}}.
\end{equation}
Similarly the Fourier coefficients of a $\Lambda$-periodic functions such as the  local volume   fulfills
\begin{equation}\label{eq:Fourier_periodic2}
  \hat{v}(\vec{k}) = V v_{\vec{Q}}
   \delta_{\vec{q},\vec{0}},
\end{equation}
with the Fourier coefficient
\begin{align}\label{eq:Fourier_periodic3}
v_{\vec{Q}} = \frac{1}{V_P} \int_{V_P} e^{i \vec{Q} \cdot \vec{x} } v(\vec{x} ) \diff \vec{x} , 
\end{align}
where the integral extends only over a primitive cell $V_P$.  The Eqs.~\eqref{eq:Fourier_periodic2},\eqref{eq:Fourier_periodic3} are the proper generalizations of Eqs.~\eqref{eq:Fourier_periodic1},\eqref{eq:def_Fourier} to arbitrary periodic modulations.   

 It  follows immediately that any nonvanishing term in the last product of 
Eq.~\eqref{eq:convolution} has to be such that
$\vec{q}''_i=-\vec{q}_i$ and $\vec{q}'_i=-\vec{q}''_i=\vec{q}_i$.
Last, some care is needed to deal with the remaining factor
$\hat{n}(\vec{k}'_0 + \vec{k}'_1 + \vec{k}'_2)$, which, upon using the decomposition of the wave vectors into  the Brillouin zone and reciprocal lattice, also can be written as $\hat{n}(\vec{Q}'_0 + \vec{Q}'_1
+ \vec{Q}'_2 + \vec{q}_0' + \vec{q}_1' + \vec{q}_2')$.  In order for this
factor to be itself nonvanishing, it is required that $\vec{q}_0' +
\vec{q}_1' + \vec{q}_2' = \vec{q}_0 +
\vec{q}_1 + \vec{q}_2$ is equal to a reciprocal lattice vector
$\vec{Q}^\sharp \in \Lambda^*$.  Therefore, one
finds
\begin{equation}
  \hat{n}(\vec{k}'_0 + \vec{k}'_1 + \vec{k}'_2) = V \sum_{\vec{Q}^\sharp\in \Lambda^*}  n_{\vec{Q}'_0 +
    \vec{Q}'_1 + \vec{Q}'_2 + \vec{Q}^\sharp}  \delta_{\vec{q}_0
    + \vec{q}_1 + \vec{q}_2,\vec{Q}^\sharp}.
 \end{equation}  
Substituting  these results into Eq.~\eqref{eq:convolution}, the sums over
vectors of the first Brillouin zone can be explicitly performed and
one finds the result
\begin{align} 
 S^{(3)}_{\vec{Q}_0\vec{Q}_1\vec{Q}_2} &
  ({\vec{q}_0,\vec{q}_{1},\vec{q}_{2}}) := S^{(3)}(\vec{k}_0,\vec{k}_1,\vec{k}_2) 
\nonumber \\
 \approx & n_{0}^2 \sum_{\vec{Q}^\sharp\in \Lambda^*} \delta_{\vec{q}_0
    + \vec{q}_1 + \vec{q}_2,\vec{Q}^\sharp} 
  \sum_{\substack{\vec{Q}'_0,\vec{Q}'_1,\vec{Q}'_2 \in
      \Lambda^* \\ \vec{Q}''_0,\vec{Q}''_1,\vec{Q}''_2 \in \Lambda^*}}
  n_{\vec{Q}'_0+\vec{Q}'_1+\vec{Q}'_2+\vec{Q}^\sharp} \nonumber \\
&  \times \prod_{i=0}^2 v_{-\vec{Q}'_i-\vec{Q}''_i}
  S^{(2)}_{\vec{Q}''_i\vec{Q}_i}(-\vec{q}_i,\vec{q}_i),
\end{align}
where $n_{0}=N/V$.  The overall factor $\delta_{\vec{q}_0 + \vec{q}_1
  + \vec{q}_2,\vec{Q}^\sharp}$  encodes the conservation of
the crystal momentum at the level of the triplet structure factor.  It
could have been introduced from the outset, but its deduction here
demonstrates the consistency of the convolution approximation with the
requirements of lattice periodicity. Note that $\vec{q}_0+\vec{q}_1+ \vec{q}_2 = \vec{Q}^\sharp \in \Lambda^*$ implies that $\vec{Q}^\sharp$ is either $\vec{0}$ or a reciprocal vector neighboring the origin. 

Last, let us specialize again to the case of a one-dimensional modulation and use the notation of the main text. We use the substitution rules $\vec{q}_0 \mapsto -\vec{q}, \vec{Q}_0 \mapsto -\vec{Q}_\mu$
\begin{align}
 S_{\mu,\mu_1 \mu_2} & (\vec{q},\vec{q}_1\vec{q}_2) = 
S^{(3)}(-\vec{q}-\vec{Q}_\mu,\vec{q}_1+ \vec{Q}_{\mu_1}, \vec{q}_2+ \vec{Q}_{\mu_2})  \nonumber \\
& \approx n_{0}^2 \sum_{\mu^\sharp = 0,\pm 1}  \delta_{-\vec{q} + \vec{q}_1+ \vec{q}_2, \vec{Q}_{\mu^\sharp} }  \nonumber \\
& \times \sum_{ \substack{\mu', \mu_1', \mu_2' \in \mathbb{Z} \\ \mu'', \mu_1'', \mu_2'' \in \mathbb{Z} }} 
n_{\mu' + \mu_1'+ \mu_2' + \mu^\sharp}  v_{-\mu'-\mu''} 
S_{-\mu'', -\mu}( - \vec{q})\nonumber \\
& \times v_{-\mu_1'- \mu_1''} S_{-\mu_1'', \mu_1}(\vec{q}_1)  
v_{-\mu_2'- \mu_2''} S_{-\mu_2'', \mu_2}(\vec{q}_2)  .
\end{align}
Next we observe that $S_{-\mu'', -\mu}(-\vec{q}) = S_{\mu\mu''}(\vec{q})$ and reverse the signs of the double primed summation indices $\mu_1'', \mu_2''$ as well as the sign in $\mu^\sharp$:

\begin{align} \label{eq:triple_modulated}
 S_{\mu,\mu_1 \mu_2} & (\vec{q},\vec{q}_1\vec{q}_2)     \approx n_{0}^2 \sum_{\mu^\sharp = 0,\pm 1}  \delta_{\vec{q}-\vec{Q}_{\mu^\sharp}, \vec{q}_1+ \vec{q}_2 }  \nonumber \\
& \times \sum_{ \substack{\mu', \mu_1', \mu_2' \in \mathbb{Z} \\ \mu'', \mu_1'' ,\mu_2'' \in \mathbb{Z} }} 
n_{\mu' + \mu_1'+ \mu_2' - \mu^\sharp}  v_{-\mu'-\mu''} S_{\mu,\mu'',}(\vec{q})\nonumber \\
& \times v_{-\mu_1'+ \mu_1''} S_{\mu_1'', \mu_1}(\vec{q}_1)  
v_{-\mu_2'+\mu_2''} S_{\mu_2'', \mu_2}(\vec{q}_2)  . 
\end{align}

\end{document}